\documentclass[%
preprint,
nofootinbib,
 amsmath,amssymb,
 aps,showkeys,pre
]{revtex4-2}

\usepackage{graphicx}% Include figure files
\usepackage{dcolumn}% Align table columns on decimal point
\usepackage{bm}% bold math
\usepackage{subfigure}
\begin{document}

%\preprint{APS/123-QED}

\title{Arbitrary number of thermally induced phase transitions in different universality classes in $XY$ models with higher-order terms}% Force line breaks with \\

\author{Milan \v{Z}ukovi\v{c}}
% \altaffiliation{Institute of Physics, Faculty of Science, P. J. \v{S}af\'arik University, Park Angelinum 9, 041 54 Ko\v{s}ice, Slovakia.}%Lines break automatically or can be forced with \\
\email{milan.zukovic@upjs.sk}
\affiliation{%
Institute of Physics, Faculty of Science, P. J. \v{S}af\'arik University, Park Angelinum 9, 041 54 Ko\v{s}ice, Slovakia.
}%


%\affiliation{
%\affiliation{
%\affiliation{%


\date{\today}% It is always \today, today,

\begin{abstract}
We propose generalized variants of the $XY$ model capable of exhibiting an arbitrary number of phase transitions only by varying temperature. They are constructed by supplementing the magnetic coupling with $n_t-1$ nematic terms of exponentially increasing order with the base $q=2,3,4$ and $5$, and increasing interaction strength. It is found that for $q=2,3$ and $4$ with sufficiently large coupling strength of the final term, the models exhibit a number of phase transitions equal to the number of the terms in the generalized Hamiltonian. Starting from the paramagnetic phase, the system transitions through the cascade of $n_t-1$ nematic phases of the orders $q^{k}$, $k=n_t-1,n_t-2,\hdots,1$, that are characterized by $q^{k}$ preferential spin directions symmetrically disposed around the circle, to the ferromagnetic (FM) phase at the lowest temperatures. Besides the BKT transition from the paramagnetic phase, all the remaining transitions have a non-BKT nature: depending on the value of $q$ they belong to either the Ising ($q=2$ and $4$) or the three-states Potts ($q=3$) universality class. For $q=5$, due to the interplay between different terms, the phase transitions between the ordered phases observed for $q<5$ split into two and the number of the ordered phases increases to $2n_t-1$. These phases are characterized by a domain structure with the gradually increasing short-range FM ordering within domains that extends to different kinds of FM ordering in the last two low-temperature phases. The respective transitions do not seem to obey any universality. 
\end{abstract}


\maketitle

%\tableofcontents

\section{\label{sec:intro}Introduction}

The Mermin--Wagner theorem~\cite{merm66} asserts that the continuous symmetry in a two-dimensional (2D) $XY$ model with nearest-neighbor interactions cannot be broken, thereby preventing a standard phase transition from occurring. However, despite this, the model is known to exhibit a Berezinskii--Kosterlitz--Thouless (BKT) phase transition~\cite{bere71,kost73} due to the presence of topological excitations known as vortices and antivortices. In the disordered phase at high temperatures, free vortices dominate, and the spin-spin correlation function exhibits an exponential decay with distance. At the BKT transition temperature ($T_{\rm BKT}$), vortices and antivortices pair up, leading to an algebraic decay of the correlation function within the quasi-long-range-ordered (QLRO) BKT phase. Unlike the typical LRO phases, the BKT phase remains critical for all temperatures below $T_{\rm BKT}$. It is worth mentioning that a related non-standard critical phase can also be observed in a 3D bulk system. In particular, the surface of such a system with sufficiently strong interactions can enter a so-called extraordinary-log surface phase, characterized by correlations that decay as a power of logarithm of the separation, i.e., much more slowly than in the BKT phase of a 2D system (see, e.g. ~\cite{metli22}). The extraordinary-log phase can be thought of as what happens when a BKT-like mechanism is pushed one dimension higher: In 2D, Goldstone modes of O(2) are just enough to destabilize true LRO but maintain algebraic order, while in 3D, when confined to a 2D surface, they are only marginally irrelevant, producing logarithmic correlations.

The standard $XY$ model can be expanded by incorporating (pseudo)nematic higher-order coupling terms, resulting in additional fractional vortex excitations. For instance, the addition of a second-order nematic term promotes the formation of half-vortices with a winding of $\pi$ in the order-parameter phase, differing from the initial $2\pi$ winding of integer vortices~\cite{lee85,kors85,carme87,carp89,shi11,hubs13,qi13,nui18,samlod24}. These half-vortices can pair up at low temperatures, connected by domain-wall strings with a finite string tension. When the nematic interaction features a positive integer $q \geq 2$ with a periodicity of $2\pi/q$, it generates fractional vortices with a noninteger ($1/q$) winding number. The interplay between integer and fractional vortices leads to a more complex critical behavior in these generalized $XY$ models. Additionally, these models have practical applications in various experimental systems, such as liquid crystals~\cite{lee85,geng09}, superconductors~\cite{hlub08}, DNA packing~\cite{grason08}, and more~\cite{bonnes12,bhas12,forg16,cairns16,zuko16}.

A well-studied model includes magnetic ($J_1$) and second-order nematic ($J_2$) terms, which exhibit a phase separation behavior at different temperatures if the relative strength of the nematic coupling is sufficiently large~\cite{lee85,kors85,carp89,shi11,hubs13,qi13,nui18,samlod24}. The high-temperature phase transition to the paramagnetic phase belongs to the BKT universality class, while the nematic-magnetic phase transition is of Ising character. Interestingly, increasing the order of the nematic term leads to a significant alteration in the critical behavior, resulting in multiple ordered phases and transitions belonging to different universality classes~\cite{pode11,cano14,cano16,nui23}. Other generalizations of the $XY$ model involve the exponential decrease of coupling terms as the order increases, the presence of which can significantly influence the low-temperature critical behavior as well as the order of the single phase transition to the paramagnetic phase~\cite{zuko17}.

The present research explores constructing generalized $XY$ models capable of showcasing a greater number of phase transitions by modifying the model Hamiltonian systematically. By adjusting temperature alone, the proposed models can exhibit numerous phase transitions, offering a controlled approach to studying complex phase diagrams. This study is motivated by the possibility of creating models that would show different numbers of phase transitions solely based on temperature variations that belong to different universality classes~\cite{pode11,cano14,cano16,nui23} and extends the investigation recently reported in~\cite{zuko24}. 

\section{\label{sec:method}Model and Method}
%\subsection{Monte Carlo}

We consider a generalized form of the classical $XY$ model (or planar rotator) with $n_t$ terms. Besides the standard bilinear coupling the Hamiltonian includes $n_t-1$ higher-order (pseudo-nematic) terms with the exponentially increasing order in the form
\begin{equation}
\label{Hamiltonian}
{\mathcal H}=-\sum_{k=0}^{n_t-1}J_{q^k}\sum_{\langle i,j \rangle}\cos(q^k\phi_{i,j}),
\end{equation}
where $\phi_{i,j}=\phi_{i}-\phi_{j}$ is an angle between the neighboring spins $i$ and $j$ and $q$ is the parameter that can take integer values larger or equal to two. The second summation runs over all nearest-neighbor pairs of spins on the lattice and the first one over all the terms in the Hamiltonian. To achieve multiple phase transitions the coupling constants $J_{q^k}$ must increase with $k$. As will be shown below, a simple linear increase of the couplings (with a small modification of the value of the highest-order term $J_{q^{n_t-1}}$) typically leads to at least $n_t$ phase transitions at different temperatures. Below, we restrict our investigation to the cases with the parameter values $n_t > 2$ and $2 \leq q \leq 5$. For better comparison in all the cases the value of the largest coupling is set to $J_{q^{n_t-1}}=k_B=1$ to fix the temperature scale.

Spin systems arranged on a square lattice of side length $L$ with periodic boundary conditions are simulated using the Metropolis algorithm. To ensure thermal averaging, we perform $2 \times 10^5$ Monte Carlo sweeps (MCS) after an initial $4 \times 10^4$ sweeps are discarded to allow the system to reach equilibrium. The temperature dependence of various thermodynamic quantities is studied by gradually cooling the system from a high-temperature paramagnetic phase, with temperature $T$, down to lower temperatures in steps of $\Delta T = 0.025$. Each simulation at a new temperature is initialized using the final configuration from the previous temperature.

To determine critical exponents and identify the universality classes of phase transitions, we carry out finite-size scaling (FSS) analysis. This involves reweighting techniques~\cite{ferr88,ferr89} applied to lattice sizes ranging from $L = 24$ to $120$. Due to the significant increase in the integrated autocorrelation time near the critical point (reaching up to $\sim 10^4$ MCS for the largest systems) we increase the number of MCS for reweighting to $10^7$, following a thermalization period of $2 \times 10^6$ MCS. Statistical uncertainties are assessed using the $\Gamma$-method~\cite{wolf04}.

Among the key quantities recorded and analyzed are the specific heat per spin $c$
\begin{equation}
c=\frac{\langle {\mathcal H}^{2} \rangle - \langle {\mathcal H} \rangle^{2}}{L^2T^{2}},
\label{c}
\end{equation}
the generalized magnetizations $m_k$, $k=1,2,\hdots,q^{n_t-1}$,
\begin{equation}
m_k=\langle M_{k} \rangle/L^2=\left\langle\Big|\sum_{j}\exp(ik\phi_j)\Big|\right\rangle/L^2,
\label{mk}
\end{equation}
and the corresponding susceptibilities $\chi_{k}$
\begin{equation}
\label{chi_mk}\chi_{k} = \frac{\langle M_{k}^{2} \rangle - \langle M_{k} \rangle^{2}}{L^2T},
\end{equation}
where $M_1$ represents the magnetic and $M_k$, for $k>1$, the (pseudo)nematic order parameters. We note that the generalized magnetizations are not true LRO but rather local order parameters for the present models. Nevertheless, they are useful and are often used in MC studies for distinguishing between different phases and for constructing approximate phase diagrams. Furthermore, we calculate the derivatives of the generalized magnetizations
\begin{equation}
\label{dm}dm_{k} = \frac{\partial}{\partial \beta}\langle M_{k} \rangle = \langle M_{k} {\mathcal H}\rangle -\langle M_{k} \rangle\langle {\mathcal H} \rangle,
\end{equation}
and the derivatives of their logarithms
\begin{equation}
\label{dlnm}dlnm_{k} = \frac{\partial}{\partial \beta}\ln\langle M_{k} \rangle = \frac{\langle M_{k} {\mathcal H}\rangle}{\langle M_{k} \rangle}- \langle {\mathcal H} \rangle.
\end{equation}

At second-order phase transitions the maxima of the above quantities scale with the lattice size as
\begin{equation}
\label{fss_c}c_{max}(L) \propto L^{\alpha/\nu},
\end{equation}
\begin{equation}
\label{fss_chi}\chi_{k,max}(L) \propto L^{\gamma/\nu},
\end{equation}
\begin{equation}
\label{fss_dmk}dm_{k,max}(L) \propto L^{(1-\beta)/\nu},
\end{equation}
\begin{equation}
\label{fss_dlnmk}dlnm_{k,max}(L) \propto L^{1/\nu},
\end{equation}
where $\alpha,\beta,\gamma$ and $\nu$ represent the critical exponents of the specific heat, the order parameter $m_k$, the susceptibility $\chi_k$ and the correlation length, respectively. 

At the BKT transition the critical exponent of the algebraically decaying correlation function  $\eta$ can be obtained from FSS of the susceptibility (\ref{fss_chi}), by using the hyperscaling relation $\eta=2-\gamma/\nu$.

\section{\label{sec:results}Results}
The critical behavior of the simplest model with $n_t=2$ and $q=2$, i.e. the $J_1-J_2$ model with the bilinear and biquadratic interactions, is well understood from the earlier studies~\cite{lee85,kors85,carme87,hlub08,geng09}. In the exchange parameter region of our interest, namely for $J_2 = 1-J_1 \gtrsim 0.675$, there are two separate phase transitions. The high-temperature one corresponds to the BKT phase transition from the paramagnetic (P) to the nematic phase, in which the spin axes align in the same direction. On the other hand, the low-temperature phase transition is associated with the flipping of a portion of spins in the opposite direction to align their heads with the majority of the spins into the ferromagnetic (FM) arrangement. Recently we have shown~\cite{zuko24} that by further generalization of this model that includes an arbitrary number of higher-order terms with the exponentially increasing order in the form~(\ref{Hamiltonian}) with $q=2$ and linearly increasing coupling constants the model can shown the number of phase transitions equal to the number of the terms in the Hamiltonian. Below, we study the critical properties of this model for $n_t > 2$ and $2 \leq q \leq 5$ and focus on the aspects of the overall number of the phase transitions, their universality classes and the nature of the observed phases.

\subsection{\label{subsec:results_nt3}Case $n_t=3$}

\begin{figure}[t!]
\centering
\subfigure{\includegraphics[scale=0.4,clip]{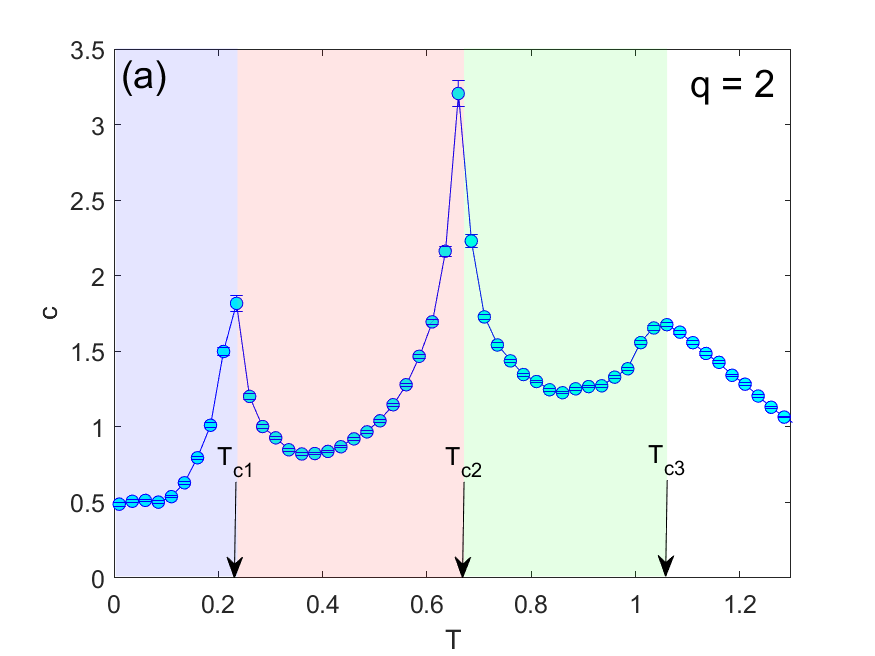}\label{fig:c-T_q2_k3_J1_01}}\hspace*{-2mm}
\subfigure{\includegraphics[scale=0.4,clip]{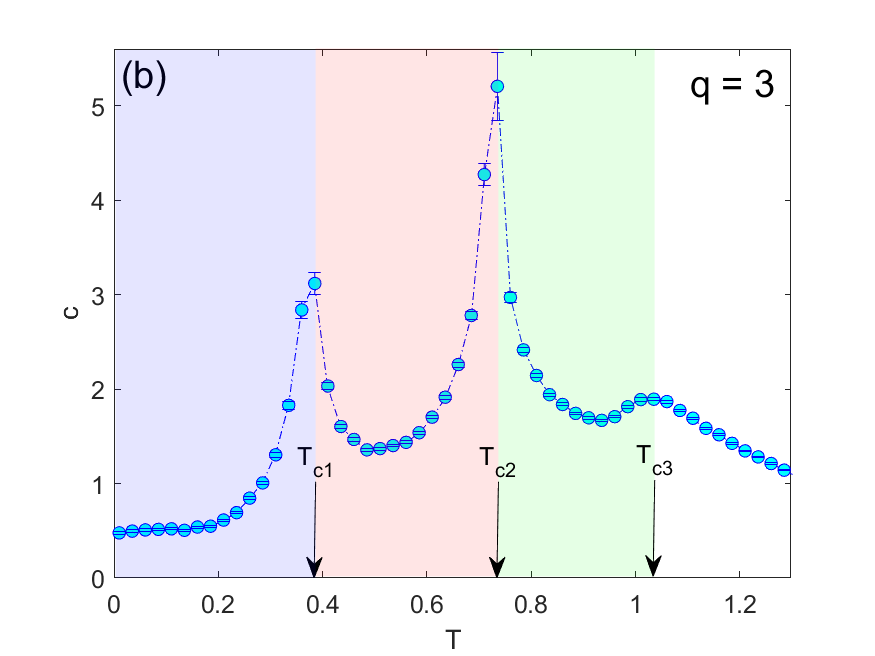}\label{fig:c-T_q3_k3_Jincr}}\hspace*{-2mm}
\subfigure{\includegraphics[scale=0.4,clip]{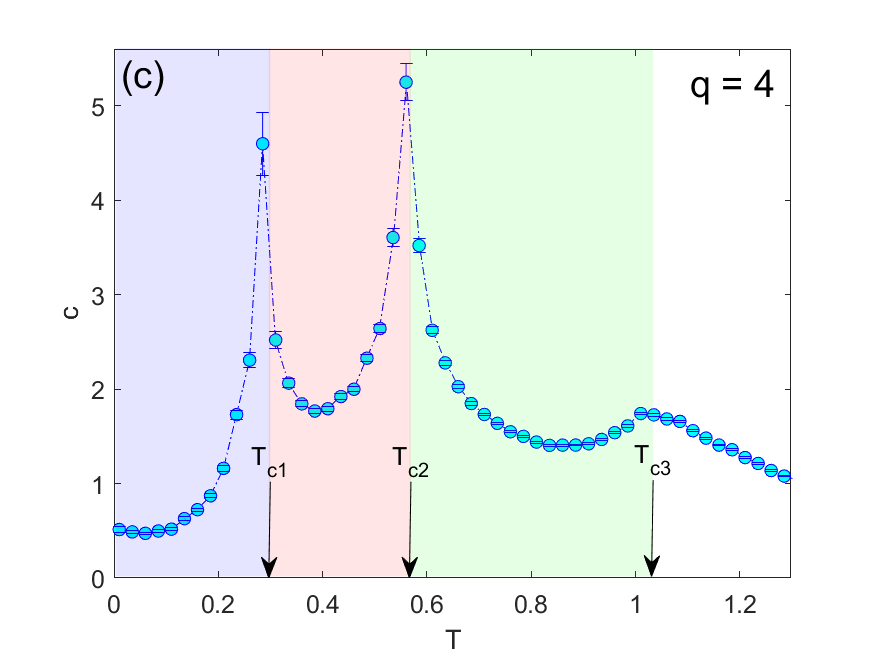}\label{fig:c-T_q4_k3_Jincr}}\\
\subfigure{\includegraphics[scale=0.4,clip]{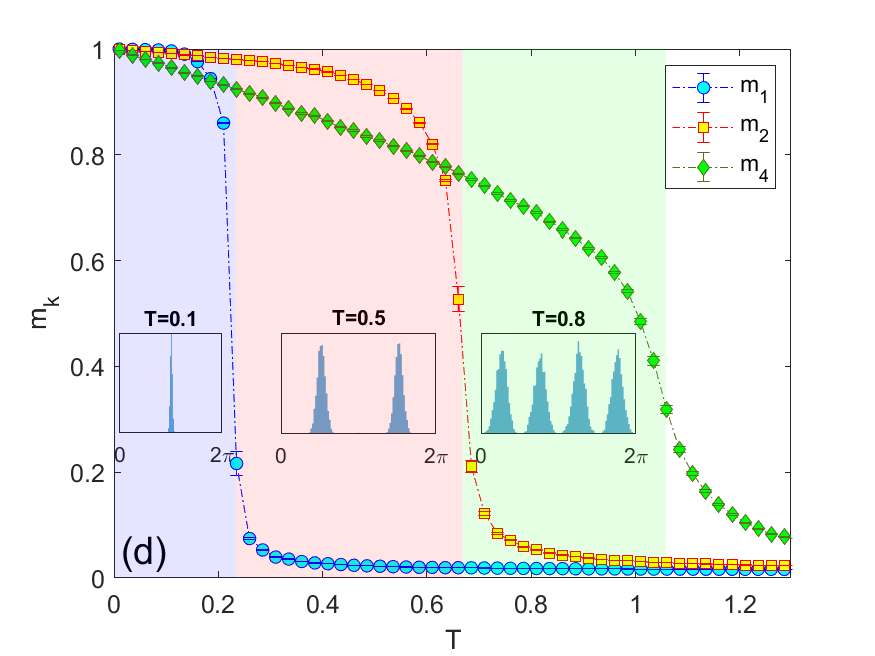}\label{fig:mk-T_q2_k3_J1_01}}\hspace*{-2mm}
\subfigure{\includegraphics[scale=0.4,clip]{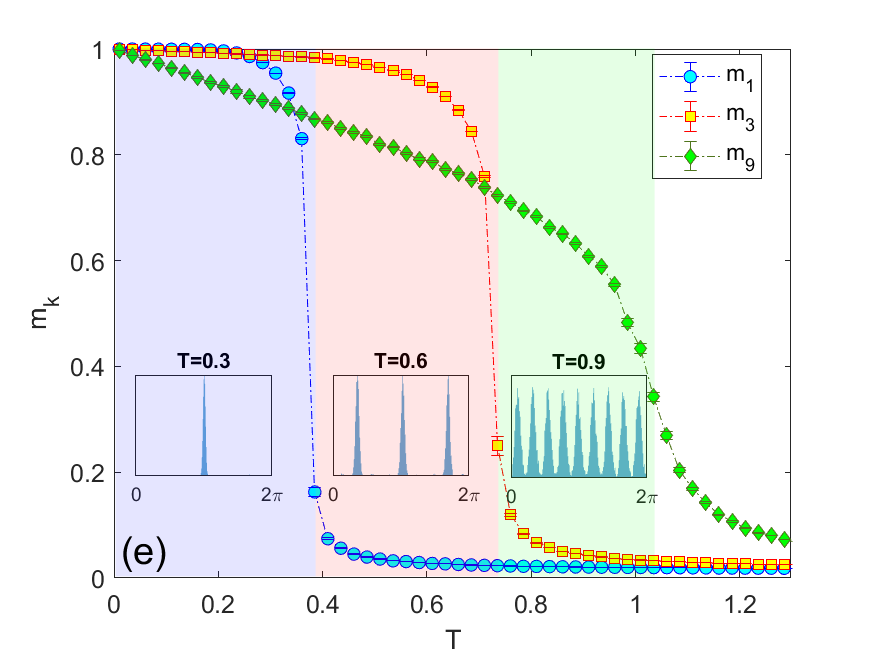}\label{fig:mk-T_q3_k3_Jincr}}\hspace*{-2mm}
\subfigure{\includegraphics[scale=0.4,clip]{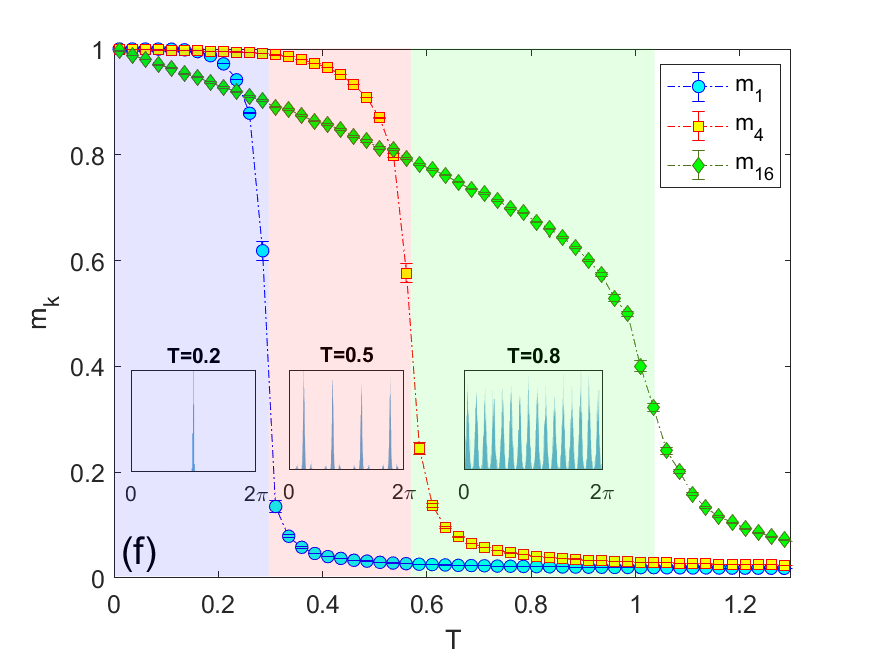}\label{fig:mk-T_q4_k3_Jincr}}\\
\subfigure{\includegraphics[scale=0.4,clip]{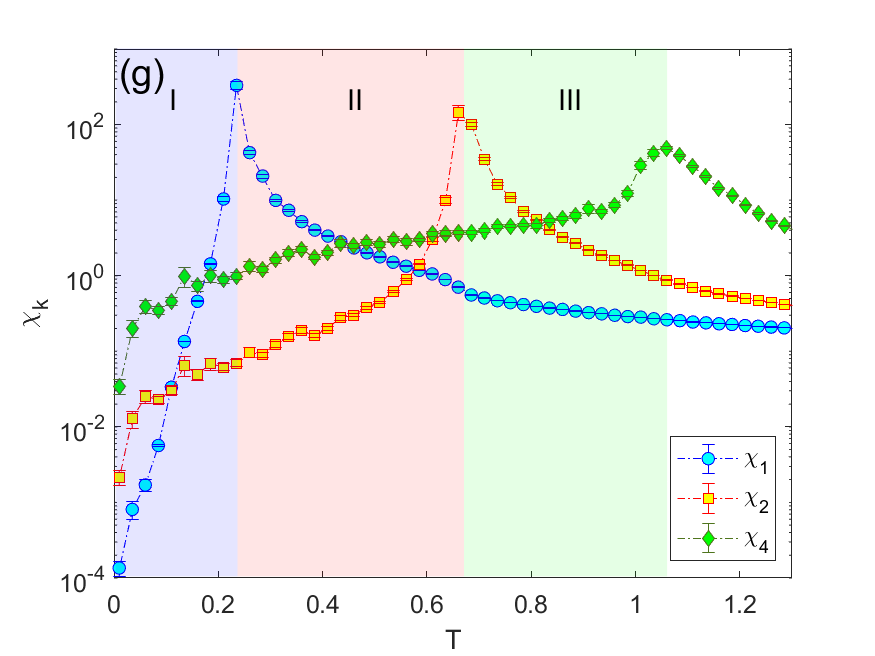}\label{fig:chik-T_q2_k3_J1_01}}\hspace*{-2mm}
\subfigure{\includegraphics[scale=0.4,clip]{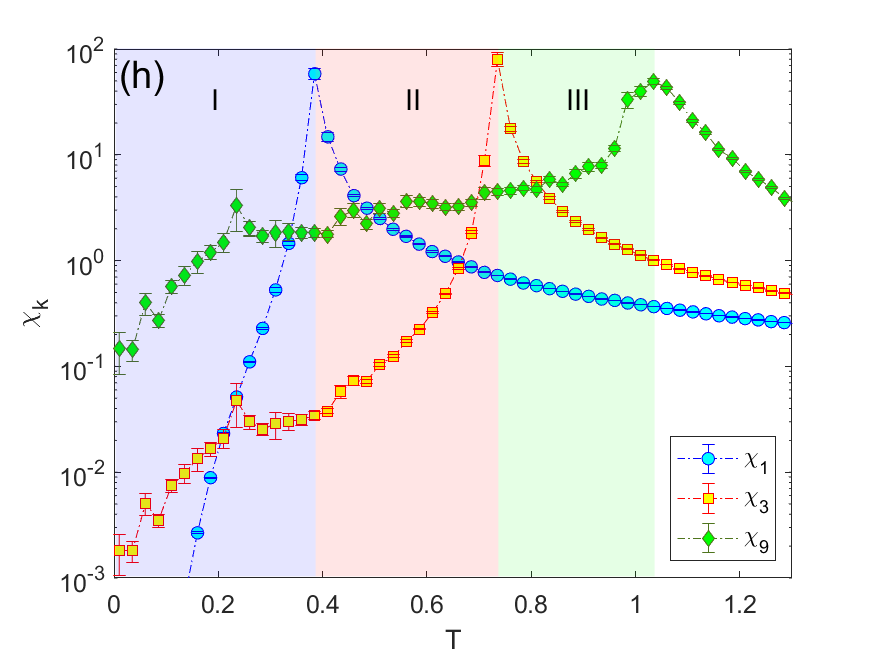}\label{fig:chik-T_q3_k3_Jincr}}\hspace*{-2mm}
\subfigure{\includegraphics[scale=0.4,clip]{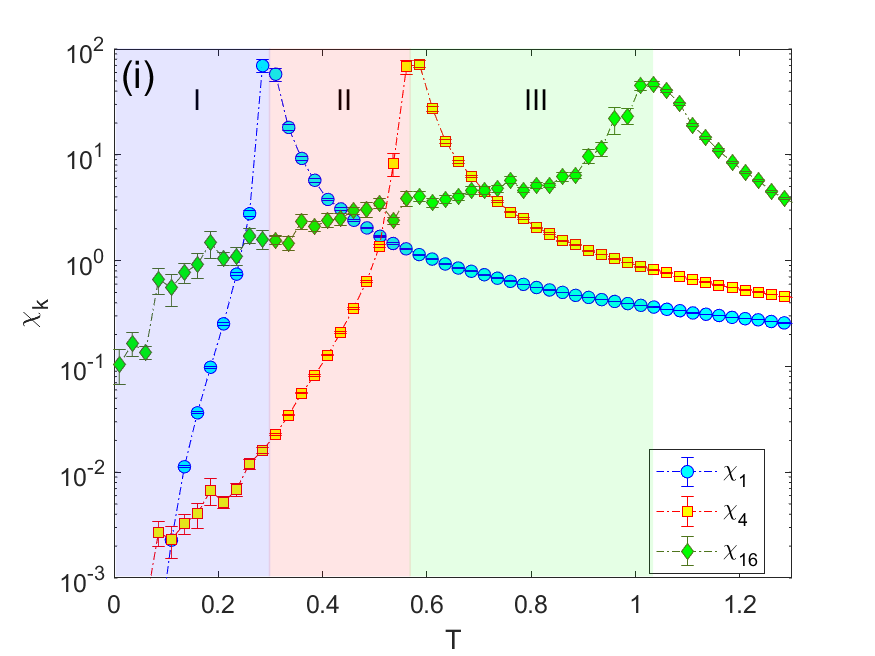}\label{fig:chik-T_q4_k3_Jincr}}
\caption{(Color online) Temperature dependencies of (a-c) the specific heat (d-f) the generalized magnetizations and (g-i) the generalized susceptibilities, for $n_t=3$ and (a,d,g) $q=2$, (b,e,h) $q=3$, and (c,f,i) $q=4$. The coupling constants are set to: $J_{1}=0.1$, $J_{2}=0.3$ and $J_{4}=1$ for $q=2$, $J_{1}=0.25$, $J_{3}=0.5$ and $J_{9}=1$ for $q=3$, and $J_{1}=0.25$, $J_{4}=0.5$ and $J_{16}=1$ for $q=4$. Background colors highlight approximate regions occupied by the phases I, II and III. Spin angle distributions at selected temperatures in the respective phases are demonstrated in the insets of panels (d-f).}\label{fig:x-T_q2-4}
\end{figure}

In Fig.~\ref{fig:x-T_q2-4} we present temperature variations of the evaluated quantities for the models with $n_t=3$, different values of $q=2,3$ and $4$, and different settings of the coupling parameters\footnote{We remark that the increasing $q$ reduces the values of observed transition temperatures below $T_{\rm BKT}$ as well as the distances between them and thus to obtain well-separated phase transitions one needs to slightly change the values of $J_k$.}. For all the cases the specific heat curves display three peaks, indicating the presence of three phase transitions. The high-temperature peaks are round and appear at roughly the same temperature that corresponds to the BKT transition from the paramagnetic state, $T_{\rm BKT}$. On the other hand, the remaining two peaks are sharp, indicating a non-BKT nature, which will be identified in the FSS analysis presented below. 

The character of the phases, separated by the transition points, can be assessed from the plots of the order parameters $m_k$. In particular, starting from low temperatures, the respective transition points are characterized by the decay of the parameters $m_{1}$ at $T_{c1}$, $m_{q}$ at $T_{c2}$ and $m_{q^2}$ at $T_{c3}$. Thus, the identified QLRO phases I, II and III (see panels with $\chi_k$) can be described by the order parameters $m_{q^k}$, for $k=0,1,$ and $2$, respectively. The spin ordering in the respective phases is demonstrated in the insets of Figs.~\ref{fig:x-T_q2-4}(d-f), in which the spin angle distributions on the lattice are presented. Approaching from the paramagnetic phase, below $T_{c3}$ (within phase III) spins align along $q^2$ preferential directions, symmetrically disposed around the circle. As the temperature is lowered in the following transitions at $T_{c2}$ and $T_{c1}$ the number of the preferential directions is geometrically reduced to $q^1$ in phase II and $q^0=1$ in phase I. Thus, by decreasing temperature the QLRO phases evolve from the pseudo-nematic phase III with the local order parameter $m_{q^2}$, through the phase II with the parameter $m_q$ to the ferromagnetic phase I with the parameter $m_1$.

The bottom row of Fig.~\ref{fig:x-T_q2-4} shows the generalizes susceptibilities $\chi_k$ that correspond to the quantities $m_k$. All of them show prominent peaks at roughly the same temperatures as the specific heat cures that correspond to the transition points $T_{c1}$, $T_{c2}$, and $T_{c3}$. One should keep in mind that, considering the QLRO nature of the phases, in the thermodynamic limit the divergent behavior of $\chi_k$ as well as vanishing of $m_k$ occur not only at the transition but at all the temperatures below the transition point.

\begin{figure}[t!]
\centering
\subfigure{\includegraphics[scale=0.4,clip]{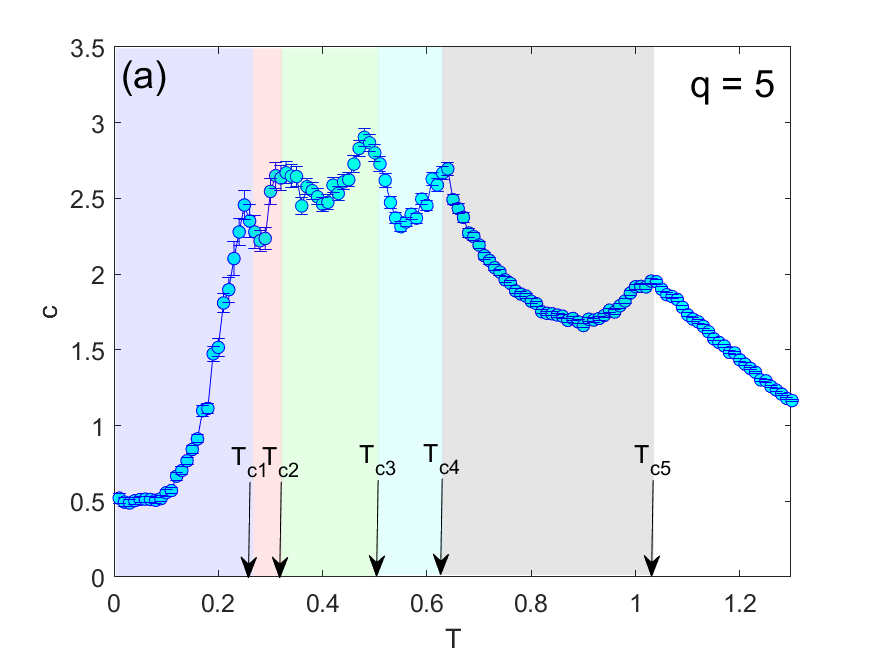}\label{fig:c-T_q5_k3_Jincr}}\hspace*{-5mm}
\subfigure{\includegraphics[scale=0.4,clip]{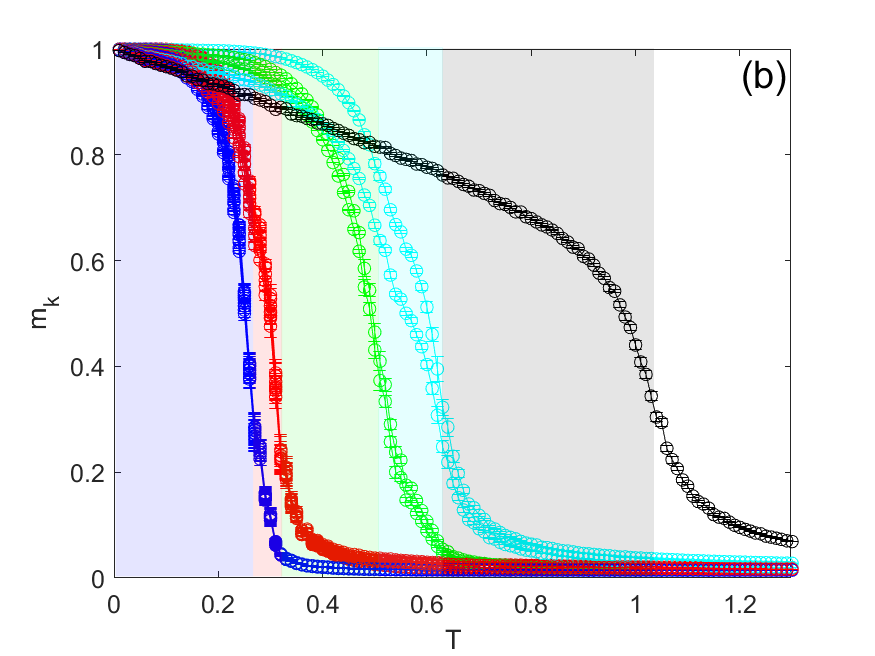}\label{fig:mk-T_q5_k3_Jincr}}\hspace*{-5mm}
\subfigure{\includegraphics[scale=0.4,clip]{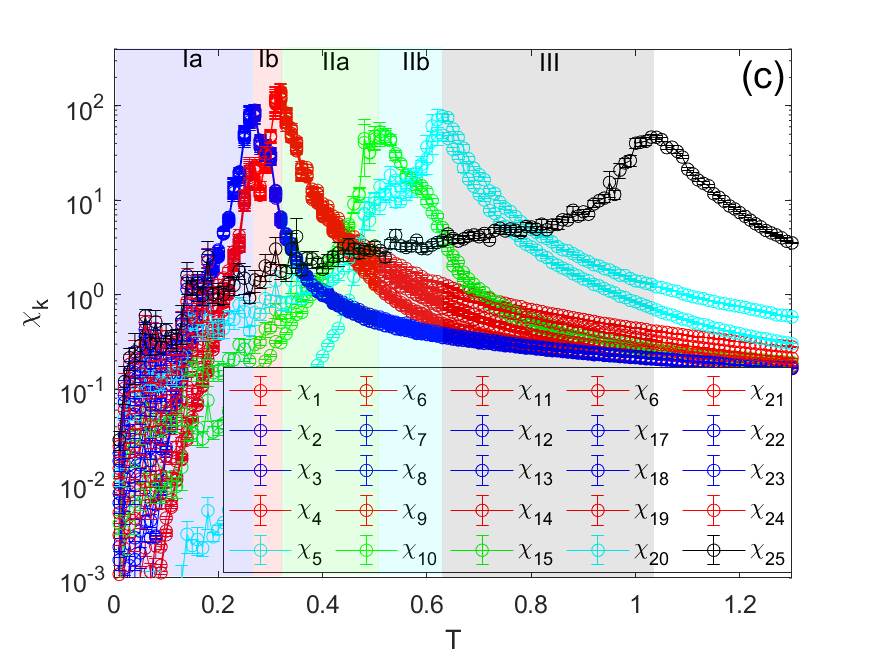}\label{fig:chik-T_q5_k3_Jincr}}
\caption{(Color online) Temperature dependencies of (a) the specific heat (b) the generalized magnetizations and (c) the generalized susceptibilities, for $n_t=3$ and $q=5$. The coupling constants are set to $J_{1}=0.3$, $J_{5}=0.6$ and $J_{25}=1$.}\label{fig:x-T_q5}
\end{figure} 

Thus, for the cases of $q=2,3$ and $4$ presented above we could observe the number of the phase transitions equal to the number of the terms in the Hamiltonian, i.e. $n_t=3$. Nevertheless, this picture changes if we continue increasing the value of the parameter $q$. In Fig.~\ref{fig:x-T_q5} we present the same quantities as in Fig.~\ref{fig:x-T_q2-4} for $q=5$. In the specific heat behavior, presented in panel (a), one can easily notice two important differences from the $q<5$ cases. First, there are apparently more than three peaks. Second, there are no prominent sharp peaks at lower temperatures. Instead, all the peaks look similar and are rather inconspicuous. 

The remaining two panels show temperature variations of $m_k$ and $\chi_k$, for all $k=1,\hdots,q^{n_t-1}$. For clarity, all the curves that decay ($m_k$) or show peak ($\chi_k$) at the same temperature are presented in the same color. One can notice that all $q^{n_t-1}=25$ curves can be classified into 5 categories (colors), which demarcate five different ordered phases. 

\begin{figure}[t!]
\centering
\subfigure{\includegraphics[scale=0.4,clip]{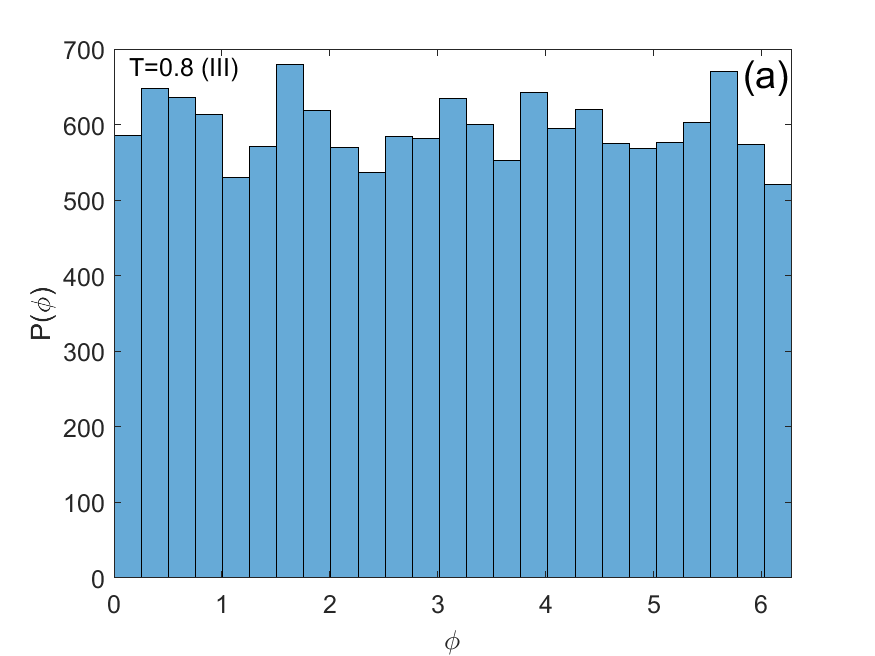}\label{fig:hst_q5_J10_3_T0_8_L120}}\hspace*{-2mm}
\subfigure{\includegraphics[scale=0.4,clip]{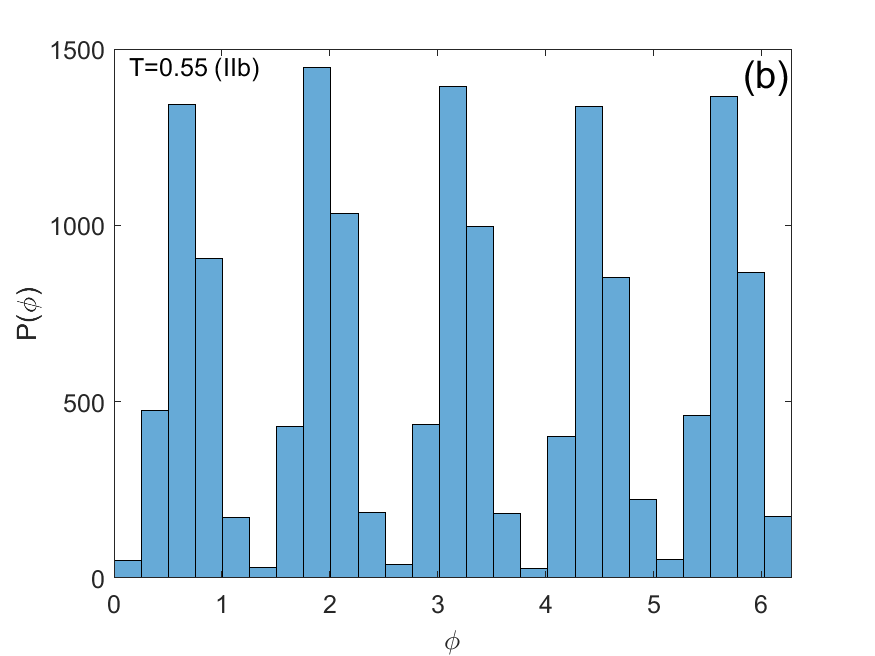}\label{fig:hst_q5_J10_3_T0_55_L120}}\hspace*{-2mm}
\subfigure{\includegraphics[scale=0.4,clip]{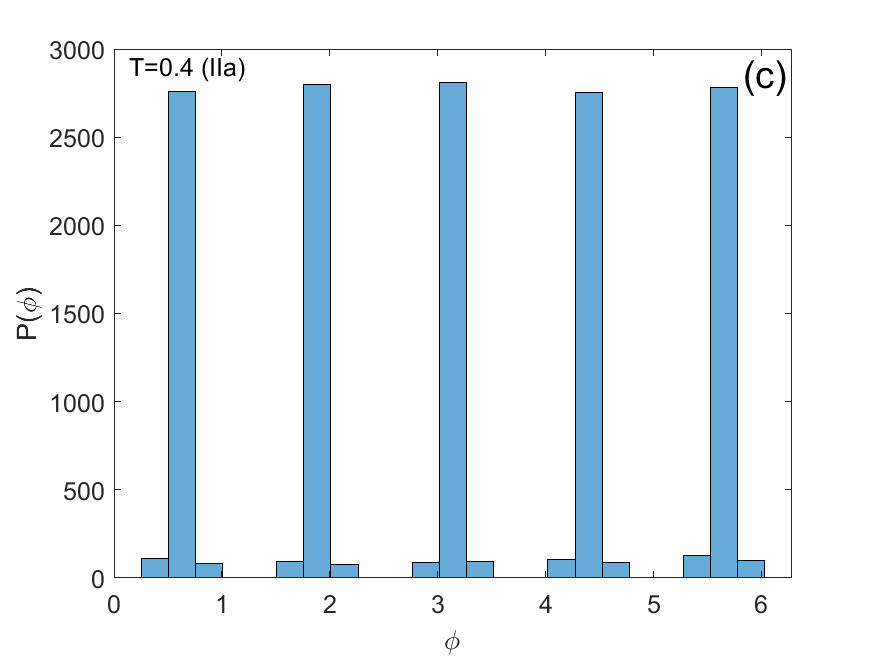}\label{fig:hst_q5_J10_3_T0_4_L120}} \\
\subfigure{\includegraphics[scale=0.4,clip]{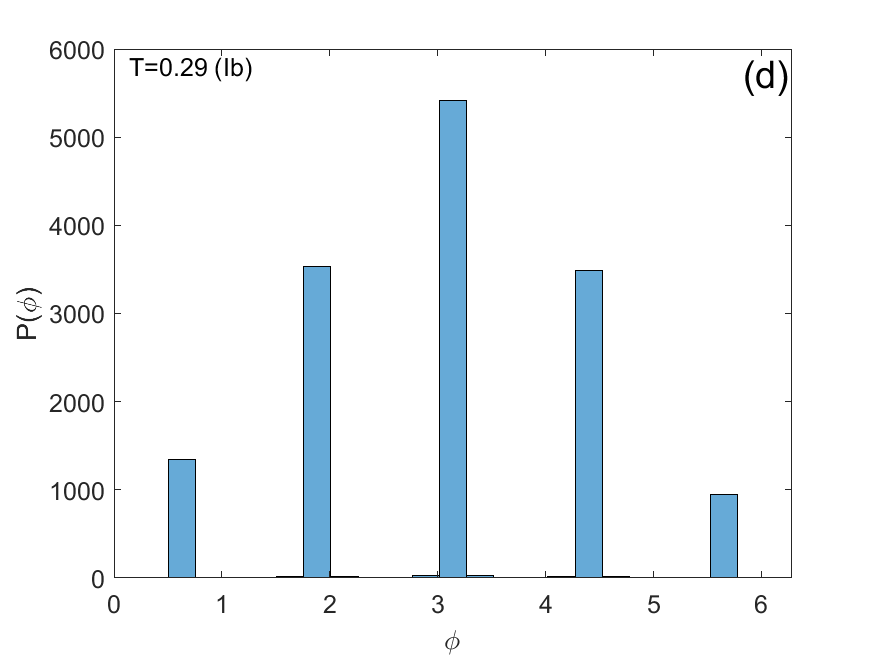}\label{fig:hst_q5_J10_3_T0_29_L120}}\hspace*{-2mm}
\subfigure{\includegraphics[scale=0.4,clip]{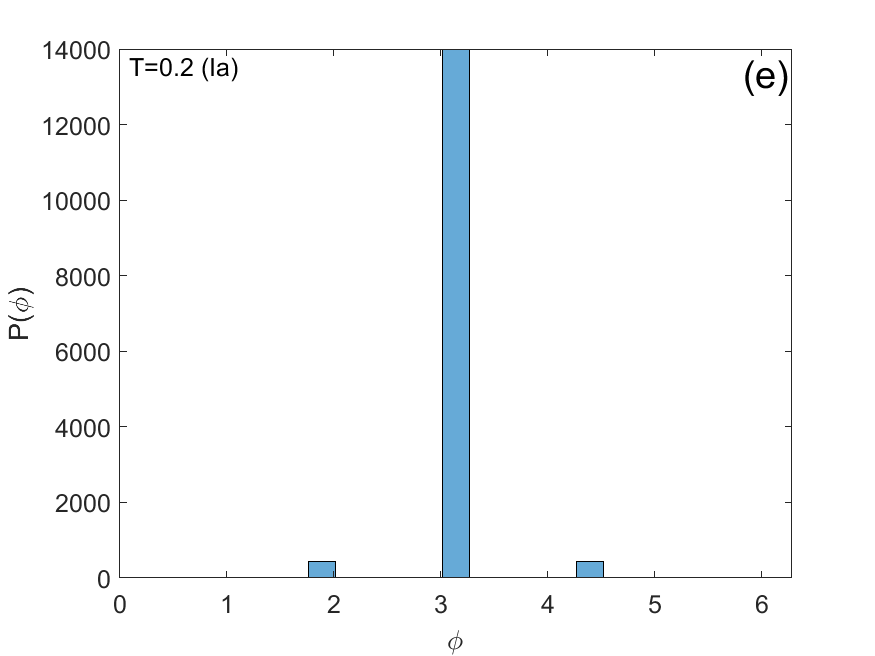}\label{fig:hst_q5_J10_3_T0_2_L120}}\\
\subfigure{\includegraphics[scale=0.4,clip]{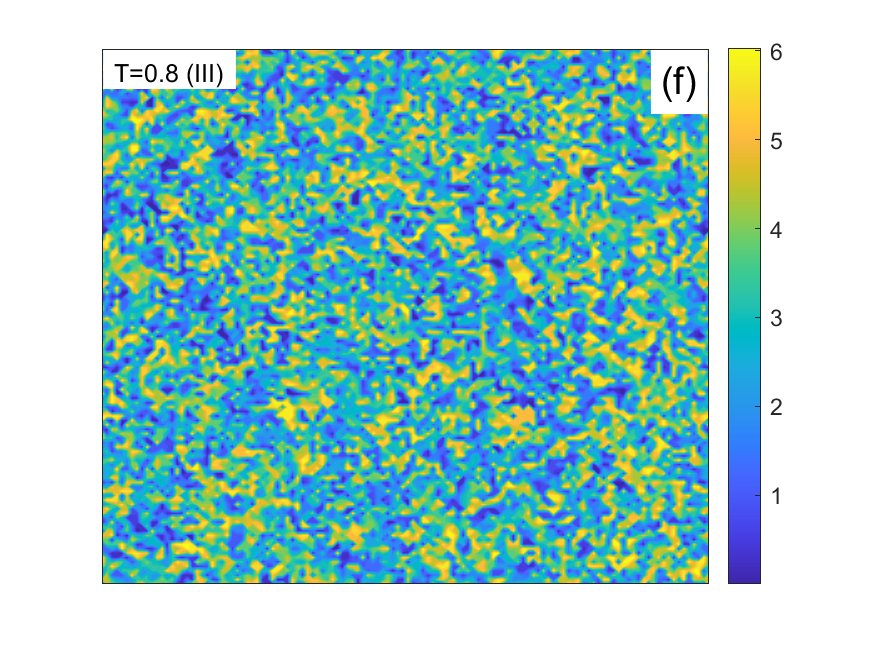}\label{fig:snp_q5_J10_3_T0_8_L120}}\hspace*{-2mm}
\subfigure{\includegraphics[scale=0.4,clip]{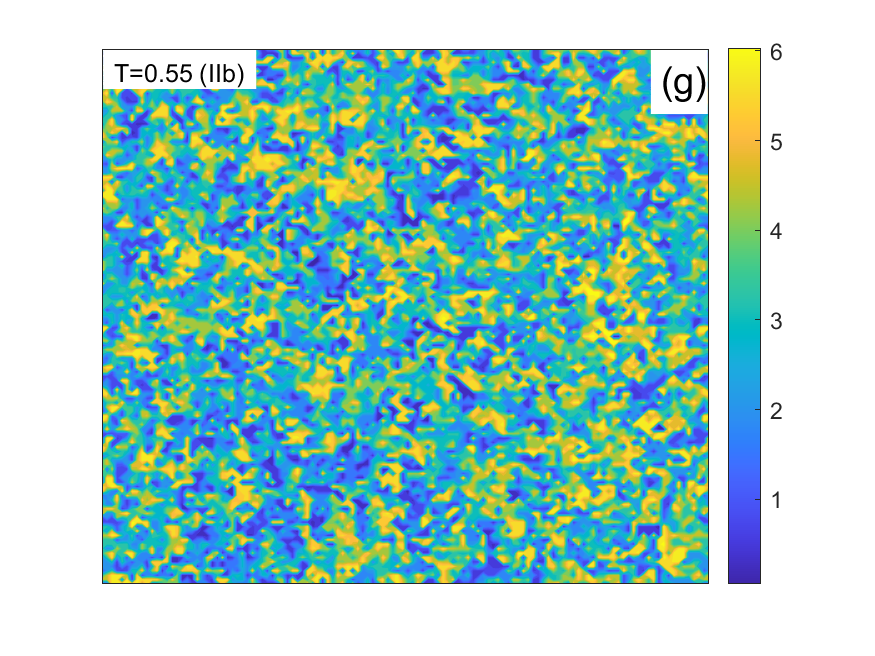}\label{fig:snp_q5_J10_3_T0_55_L120}}\hspace*{-2mm}
\subfigure{\includegraphics[scale=0.4,clip]{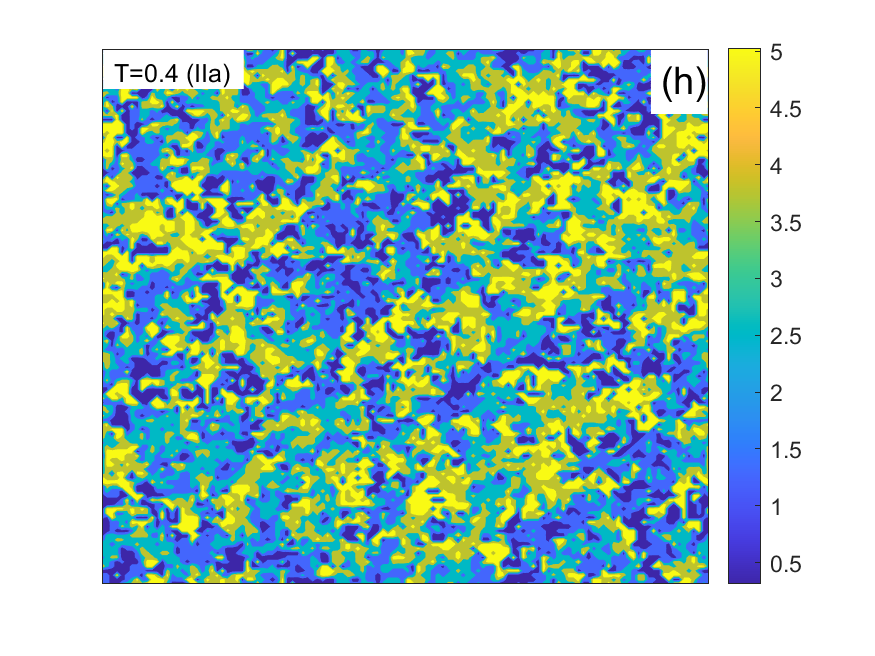}\label{fig:snp_q5_J10_3_T0_4_L120}} \\
\subfigure{\includegraphics[scale=0.4,clip]{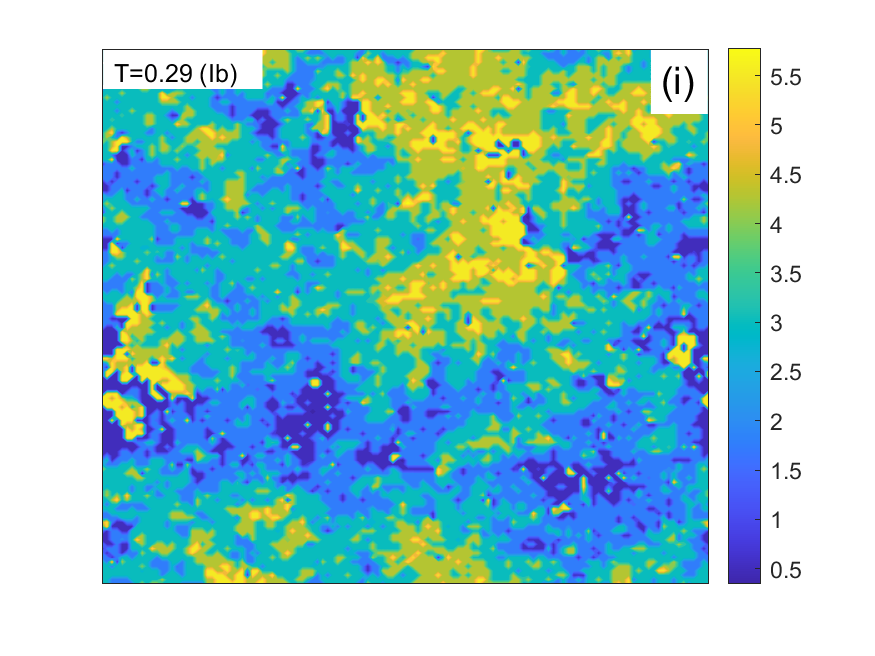}\label{fig:snp_q5_J10_3_T0_29_L120}}\hspace*{-2mm}
\subfigure{\includegraphics[scale=0.4,clip]{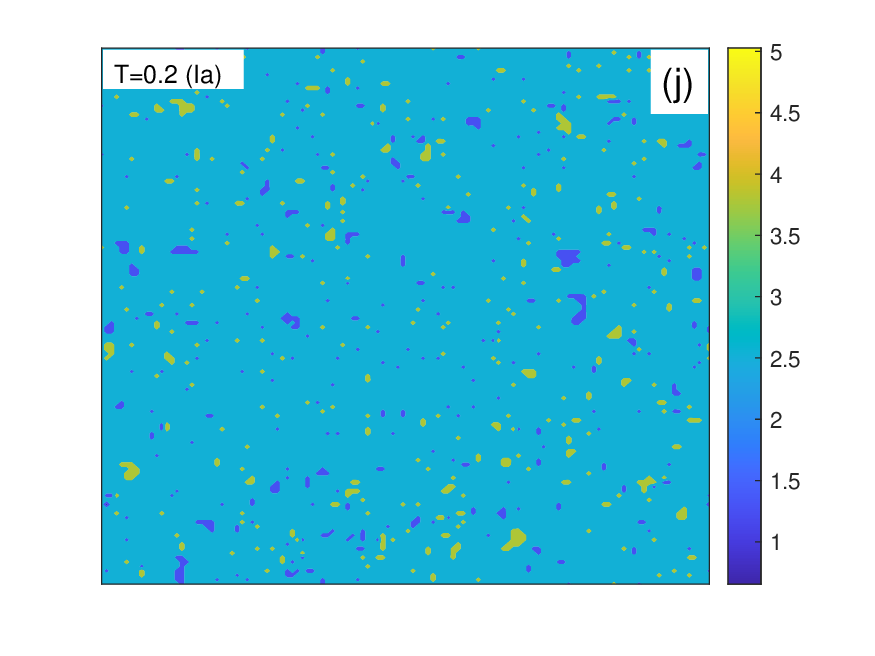}\label{fig:snp_q5_J10_2_T0_8_L120}}
\caption{(Color online) (a-e) Histograms and (f-j) real space snapshots at the representative temperatures corresponding to different phases, for $n_t=3$ and $q=5$.}\label{fig:hist_snap_nk3}
\end{figure} 

To understand the character of the respective ordered phases, in Fig.~\ref{fig:hist_snap_nk3} we show spin angle distributions (a-e) along with the corresponding real space snapshots (f-j) at the selected temperatures in different phases. For clarity, the histograms are discretized into $q^{n_t-1}=25$ bins. Apparently, the highest-temperature phase III (panels (a,f)) and the lowest-temperature phase Ia (panels (e,j)) are the same as the respective phases III and I for the cases of $q=2,3$ and $4$, presented in Fig.~\ref{fig:x-T_q2-4}. In the phase III spins align along $q^2=25$ preferential directions, symmetrically disposed around the circle, and in the phase Ia, except small thermally-driven fluctuations, all point along the same preferential direction. The phases III and Ia result from the presence of the couplings $J_{25}$ and $J_{1}$, respectively. 

\begin{figure}[t!]
\centering
\includegraphics[scale=0.6,clip]{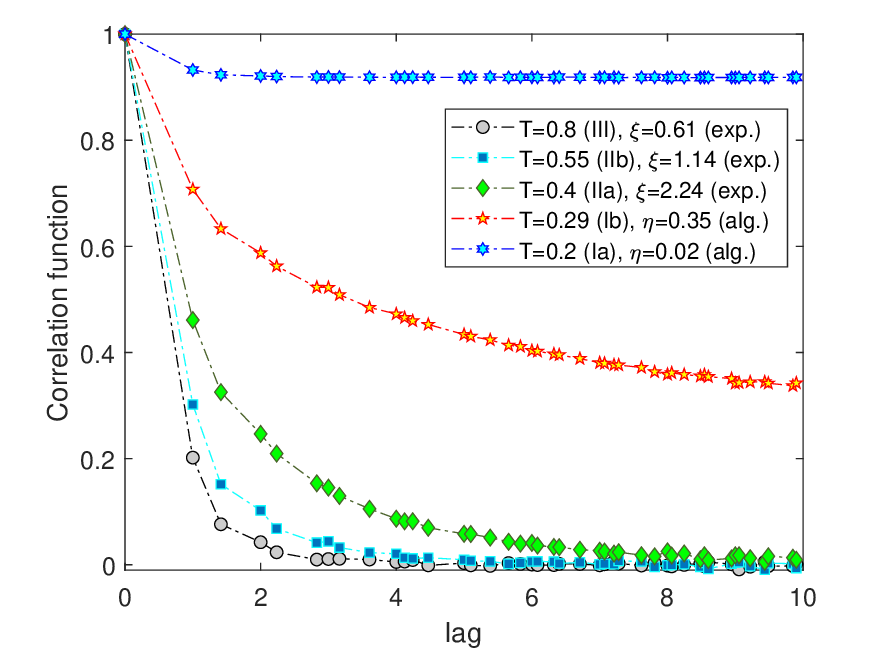}
\caption{(Color online) Spatial correlation function decay in the respective phases, represented by temperature $T$, for $n_t=3$ and $q=5$. $\xi$ denotes the estimated correlation length for the exponential (exp.) decay regime (phases IIa, IIb and III) and $\eta$ is the estimated algebraic (alg.) decay exponent (phases Ia and Ib).}\label{fig:cf_q5_nk3}
\end{figure} 

On the other hand, more interesting are the intermediate-temperature phases, which result from the interplay between all three couplings. The phase IIb emerges as a combined effect of both $J_{25}$ and $J_{5}$. The latter makes the 25 bins in the phase III split into 5 larger modes, each including 5 smaller bins. Further decrease of temperature enhances the effect of $J_5$ by making the system transition in the phase IIa, characterized with only 5 dominant well separated and equally populated preferential directions. It is worth remarking that the decreasing temperature also enhances the effect of $J_1$. Although all the phases III, IIb and IIa are of nematic nature with zero magnetic moment, there is an increasing short-range ferromagnetic ordering. The gradual increase of the ferromagnetic domains can be observed in the snapshots in panels (f-h) but also from the exponential decay of the correlation function with the gradually increasing correlation length, presented in Fig.~\ref{fig:cf_q5_nk3}.

Further decrease of temperature results in the phase Ib that pertains the 5 preferential spin ordering directions, observed in the phase IIa, however, they are not evenly populated (see panel (d)). Consequently, the total magnetic moment does not vanish and the system is composed of relatively large ferromagnetic domains (see panel (i)) with an algebraically decaying correlation function (Fig.~\ref{fig:cf_q5_nk3}). The coexistence of different domains with 5 preferential spin orientations results in the presence of the domain walls and thus the decay of the correlation function is much faster than in the lowest-temperature contiguous-domain ferromagnetic phase Ia (compare $\eta=0.35$ at $T=0.29$ and $\eta=0.02$ at $T=0.2$).

\begin{figure}[t!]
\centering
\includegraphics[scale=0.6,clip]{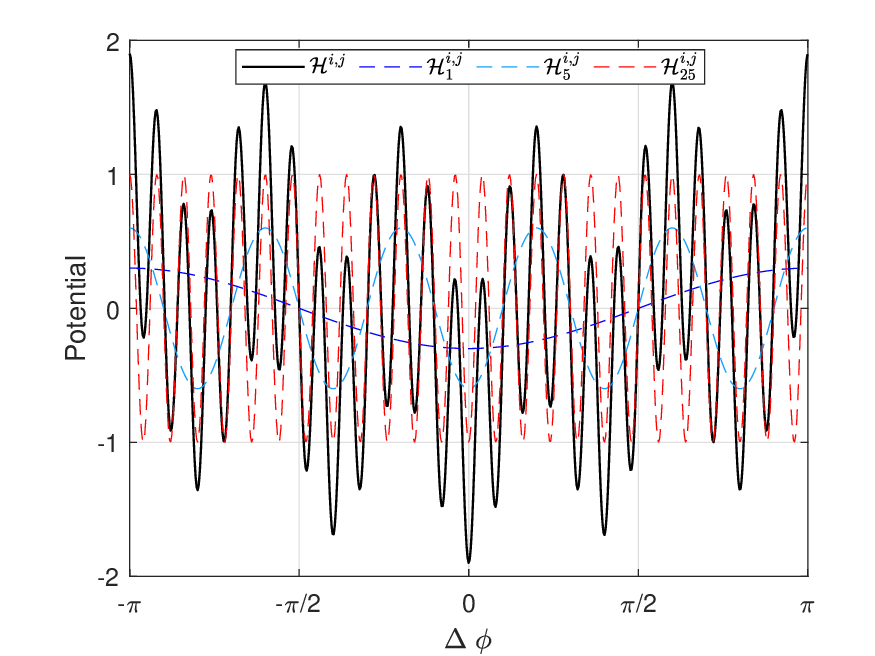}
\caption{(Color online) Potential function, ${\mathcal H^{i,j}}$, and its contributions, ${\mathcal H_{1}^{i,j}}$, ${\mathcal H_{5}^{i,j}}$ and ${\mathcal H_{25}^{i,j}}$, coming from the $n_t=3$ terms as functions of the phase difference $\Delta \phi$.}\label{fig:e_gs_q5_nk3}
\end{figure}

Apparently, the appearance of different phases can be encouraged or discouraged by the setting of the coupling parameters. In Fig.~\ref{fig:e_gs_q5_nk3} we present the potential of a pair of interacting spins, ${\mathcal H^{ij}}$, with $J_{1}=0.3$, $J_{5}=0.6$ and $J_{25}=1$, as a function of the phase difference. While the ground state energy corresponds to the $\Delta \phi = 0$ (ferromagnetic state), at higher energy values one can observe a number of local minima at different levels that due to entropic contributions may encourage non-equally weighted states with non-zero values of the phase difference. Thus, the phases like those in Fig.~\ref{fig:hist_snap_nk3} may appear due to the interplay between the potential function shape, controlled by the coupling constants setting, and the entropic contributions. 

%\subsection{\label{subsec:results_transitions}Phase transitions}

In the following, let us study the nature of the identified phase transitions. As it is well known, for $n_t=2$ and $q = 2$ ($J_1-J_2$ model), there are two transitions for $J_2 = 1-J_1 \gtrsim 0.675$: as temperature decreases there is first a BKT transition to the phase with a local nematic ordering followed by another transition to the phase with a local ferromagnetic alignment. The latter up--down symmetry-breaking transition has been confirmed to belong to the Ising universality class~\cite{lee85,kors85,carp89,shi11,hubs13,qi13,nui18,samlod24}. In the generalized $J_1-J_q$ models, for $q=3$ and $4$, it has been found that, while the high-temperature transition from the paramagnetic to the nematic phase is always of the BKT type, the low-temperature nematic-ferromagnetic transitions are either of the three-states Potts model ($q = 3$) or Ising ($q = 4$) universality class, in analogy with the discrete Clock model~\cite{cano16}. 

\begin{figure}[t!]
\centering
\subfigure{\includegraphics[scale=0.4,clip]{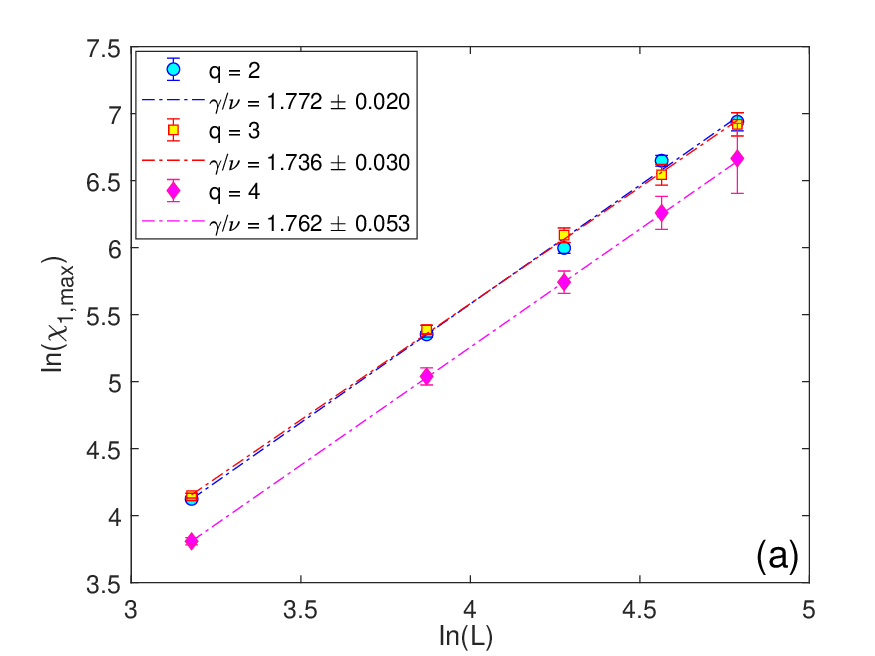}\label{fig:Tc1_q2-4_k3_Jincr_fss_xi}}\hspace*{-2mm}
\subfigure{\includegraphics[scale=0.4,clip]{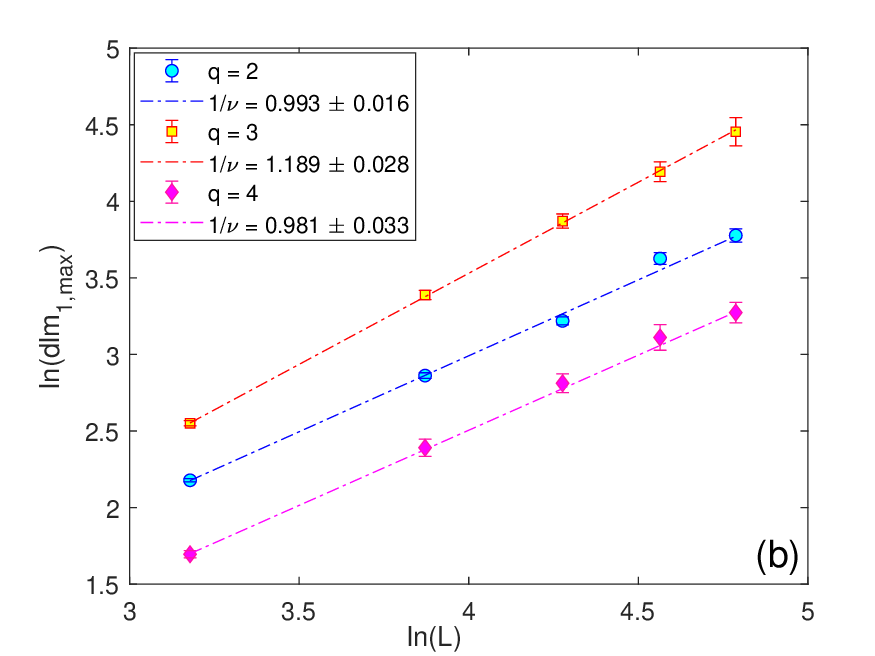}\label{fig:Tc1_q2-4_k3_Jincr_fss_dlm}}\hspace*{-2mm}
\subfigure{\includegraphics[scale=0.4,clip]{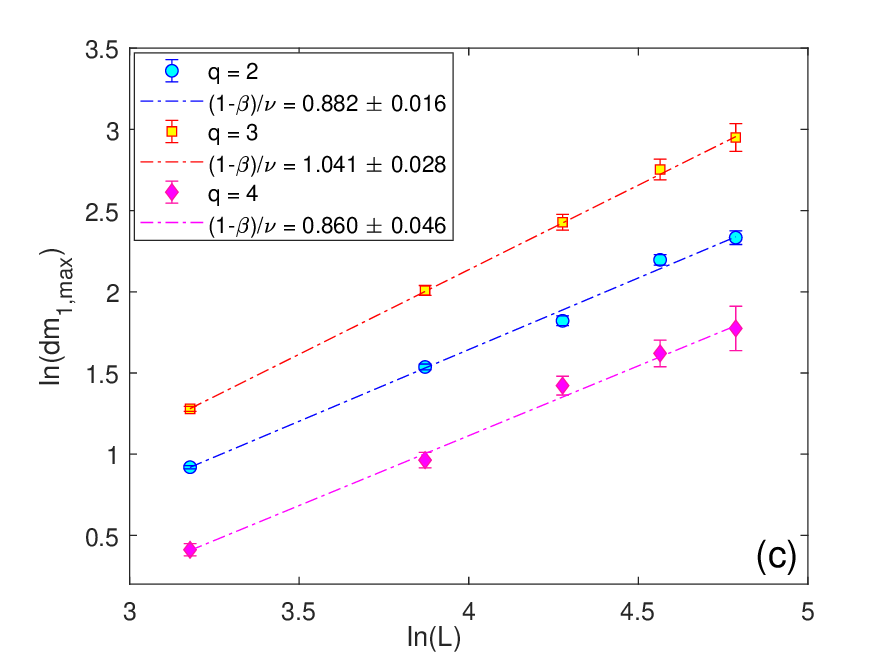}\label{fig:Tc1_q2-4_k3_Jincr_fss_dm}}\\
\subfigure{\includegraphics[scale=0.4,clip]{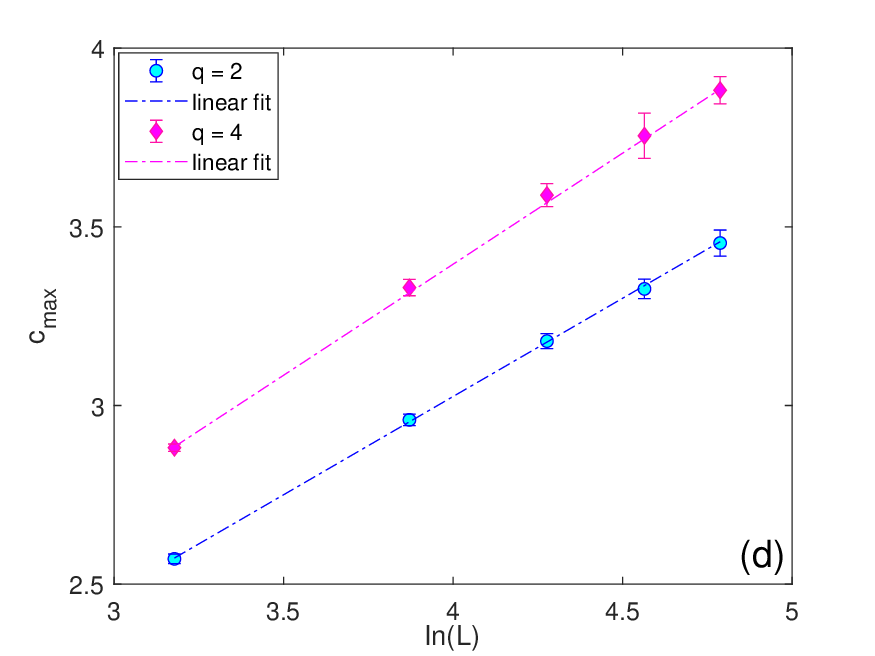}\label{fig:Tc1_q2_4_k3_Jincr_fss_c}}\hspace*{-2mm}
\subfigure{\includegraphics[scale=0.4,clip]{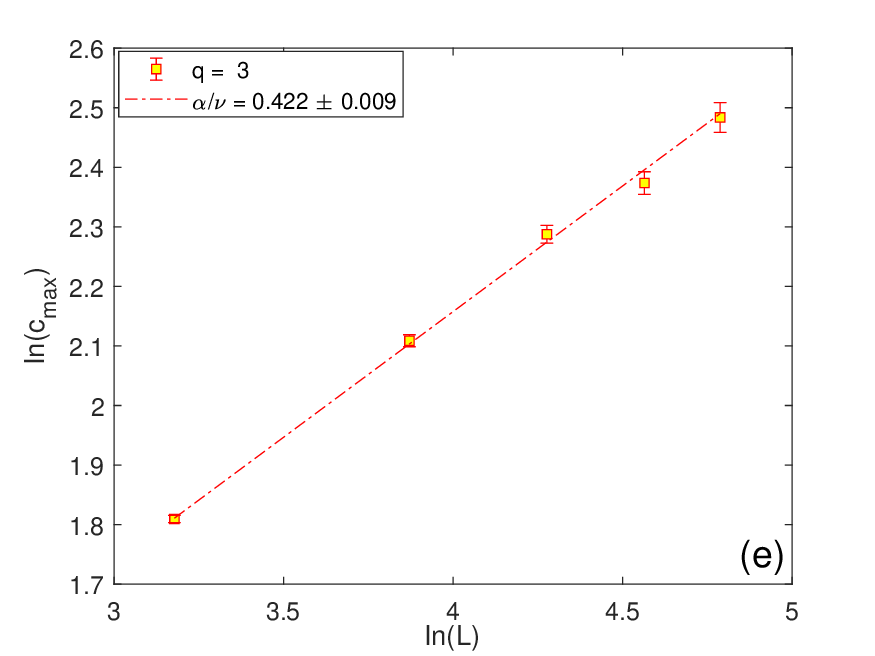}\label{fig:Tc1_q3_k3_Jincr_fss_c}}
\caption{(Color online) FSS analysis of (a) $\chi_1$, (b) $dlm_1$ (c) $dm_1$, and $c$ for (d) $q=2$ and $4$, and (e) $q=3$, with $n_t=3$ at the I-II phase transition.}\label{fig:fss_q2-4}
\end{figure}

In the present models we expect that if the additional terms in the Hamiltonian with a sufficiently large coupling $J_{q^{n_t-1}}$ trigger emergence of a new (higher-order nematic) phase then the transition from the paramagnetic phase remains BKT but the following transitions to the lower-order nematic phases that break discrete symmetry by the $q$-fold reduction of degrees of freedom, like in the $J_1-J_q$ models, should belong to the same universality, i.e. the Ising (for $q = 2$ and 4) and three-states Potts model for $q = 3$. The FFS analysis in Fig.~\ref{fig:fss_q2-4}, presented for $q = 2, 3$ and 4 at the lowest transition temperature $T_{c1}$, indeed confirms this scenario. Namely, the obtained critical exponents ratios comply with the expected Ising values $\gamma/\nu = 7/4$, $1/\nu = 1$, $(1-\beta)/\nu = 7/8$ and $\alpha/\nu = 0$ (logarithmic divergence) for $q=2$ and $4$, and the three-states Potts values $\gamma/\nu = 26/15$, $1/\nu = 6/5$, $(1-\beta)/\nu = 16/15$ and $\alpha/\nu = 2/5$ for $q=3$. The FSS at $T_{c2}$ gives similar results (not shown), confirming the theoretical expectations that both transitions below $T_{\rm BKT}$ should belong to the same universality class.

\begin{figure}[t!]
\centering
\includegraphics[scale=0.6,clip]{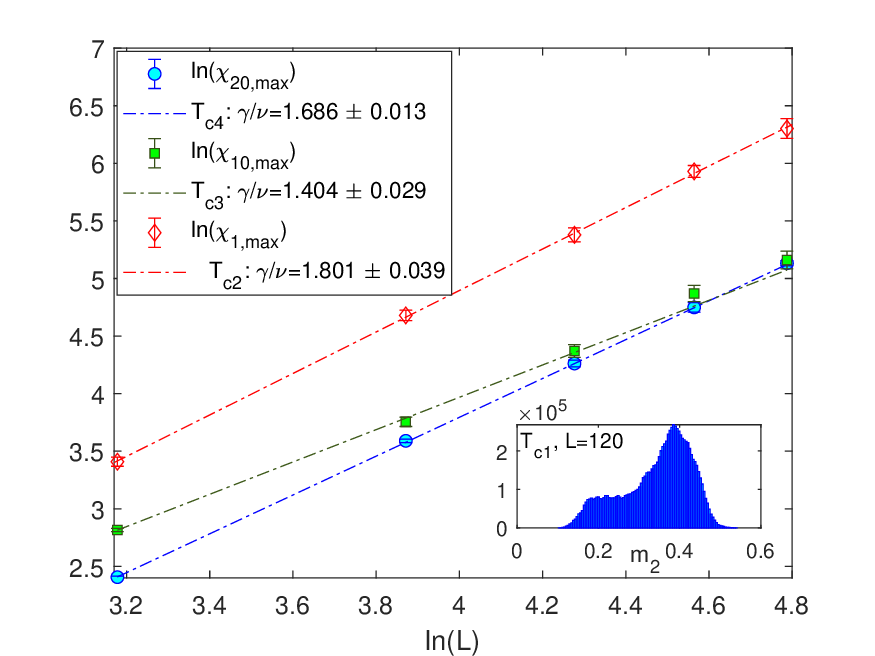}\label{fig:Tci_q5_k3_Jincr_fss_xi}
\caption{(Color online) FSS analysis of $\chi_k$, for $k=1,10$ and $20$, at the transition temperatures $T_{c2}$, $T_{c3}$ and $T_{c4}$, respectively, for $n_t=3$ and $q=5$. The inset shows the histogram of the order parameter $m_2$ at the transition temperature $T_{c1}$ for $L=120$.}\label{fig:fss_q5}
\end{figure}

On the other hand, the nature of the phase transitions observed for $q=5$ is more obscure. The FSS analysis of the generalized susceptibilities, $\chi_1$, $\chi_{10}$ and $\chi_{20}$, that correspond to the representative order parameters, $m_1$, $m_{10}$ and $m_{20}$, vanishing at the transition temperatures $T_{c2}$, $T_{c3}$ and $T_{c4}$, respectively, is presented in Fig.~\ref{fig:fss_q5}. We note that we did not manage to obtain reliable scaling behavior at the lowest transition temperature $T_{c1}$. The reason is the behavior of the corresponding order parameter $m_2$, which at larger lattice sizes shows signs of a bimodal distribution that might suggest that the transition is of first order (see the inset). However, the corresponding energy distribution does not show any such signs and thus it is more likely that the anomalous behavior of the order parameter is due to the sluggish dynamics, which becomes prominent at low temperatures and larger system sizes. Despite decreased quality of the fits, the obtained values of the critical exponents ratios at $T_{c2}$, $T_{c3}$ and $T_{c4}$ differ from each other beyond error bars. Except for $\gamma/\nu = 1.801 \pm 0.039$ at $T_{c2}$, which is compatible with the BKT value $\gamma/\nu = 2-\eta= 7/4$, the remaining values deviate from any known universality class. Nevertheless, we cannot rule out the possibility that larger system sizes and logarithmic corrections, when considered, may contribute to either restoring universality or indicating a crossover rather than a true transition.

\subsection{\label{subsec:results_nt5}Case $n_t=5$}

\begin{figure}[t!]
\centering
\subfigure{\includegraphics[scale=0.4,clip]{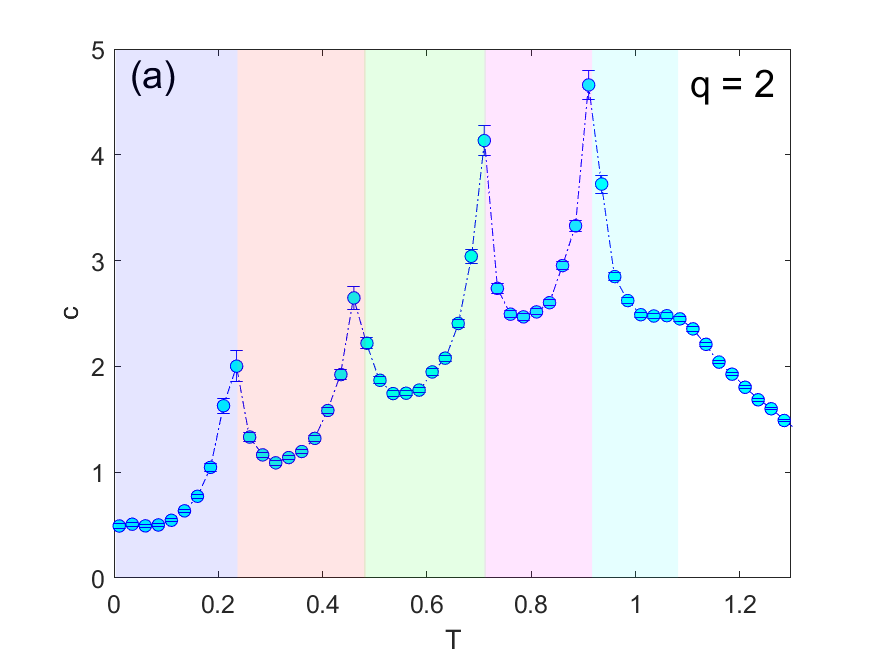}\label{fig:c-T_q2_k5_Jincr}}\hspace*{-2mm}
\subfigure{\includegraphics[scale=0.4,clip]{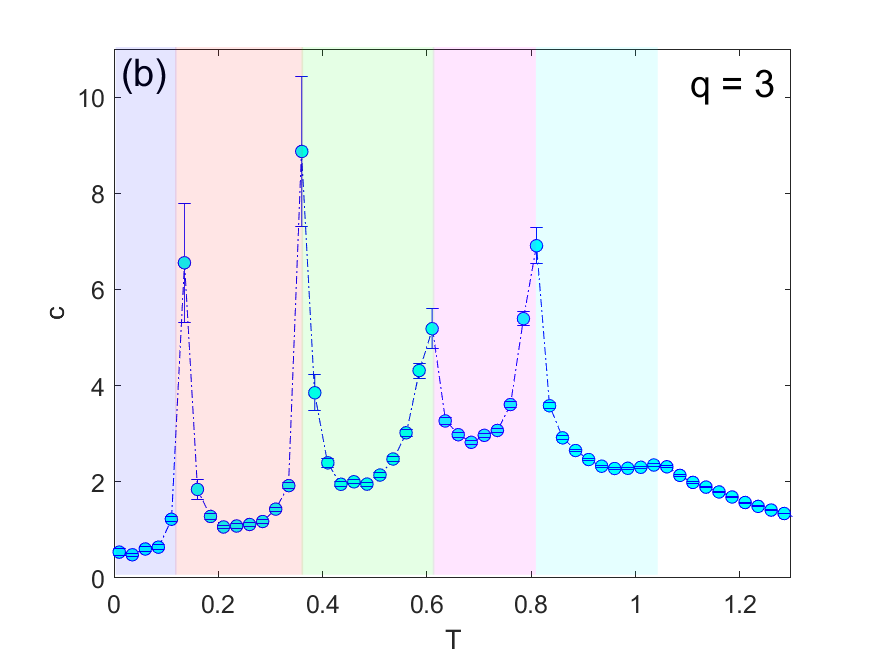}\label{fig:c-T_q3_k5_Jincr}}\hspace*{-2mm}
\subfigure{\includegraphics[scale=0.4,clip]{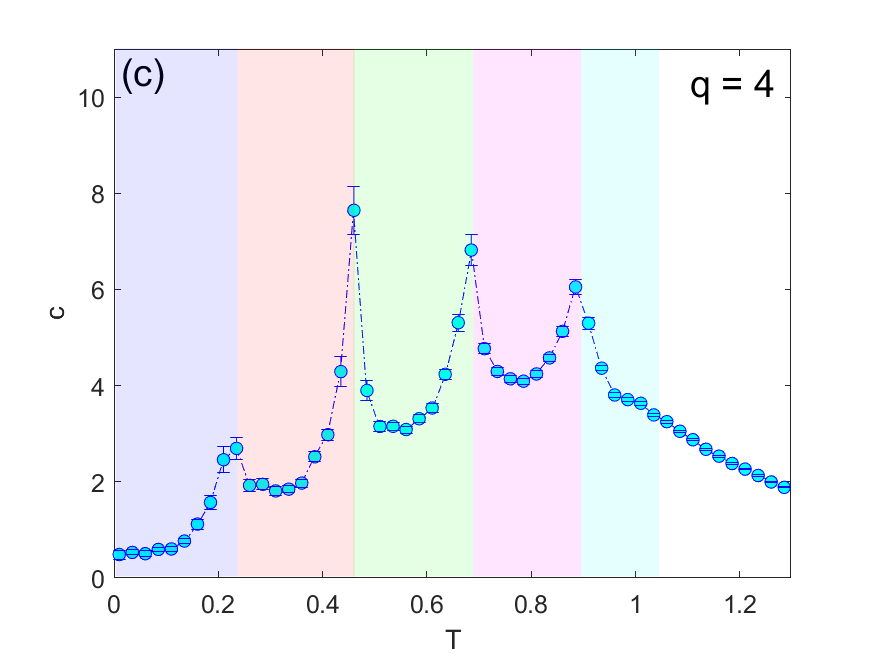}\label{fig:c-T_q4_k5_Jincr}}\\
\subfigure{\includegraphics[scale=0.4,clip]{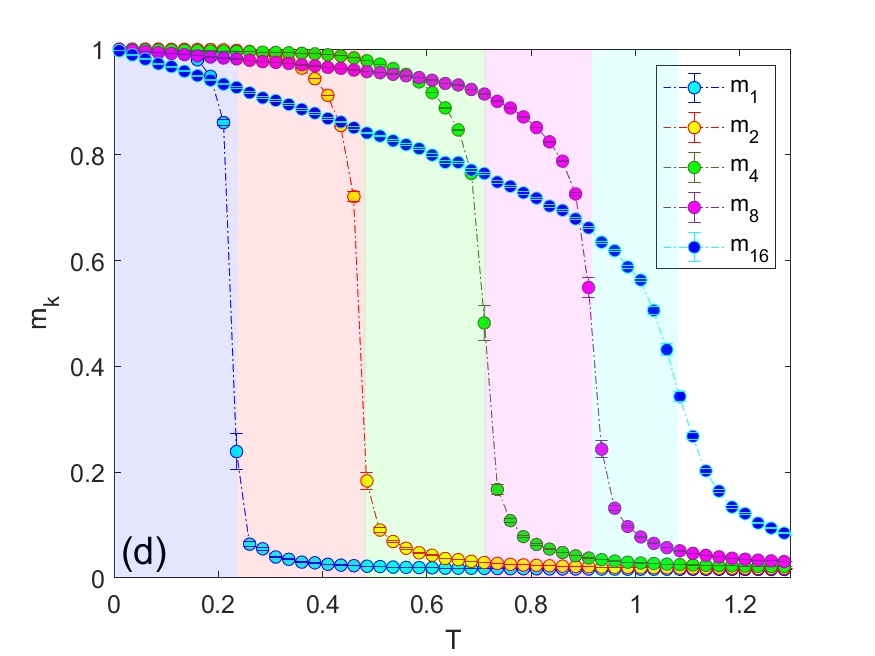}\label{fig:mk-T_q2_k5_Jincr}}\hspace*{-2mm}
\subfigure{\includegraphics[scale=0.4,clip]{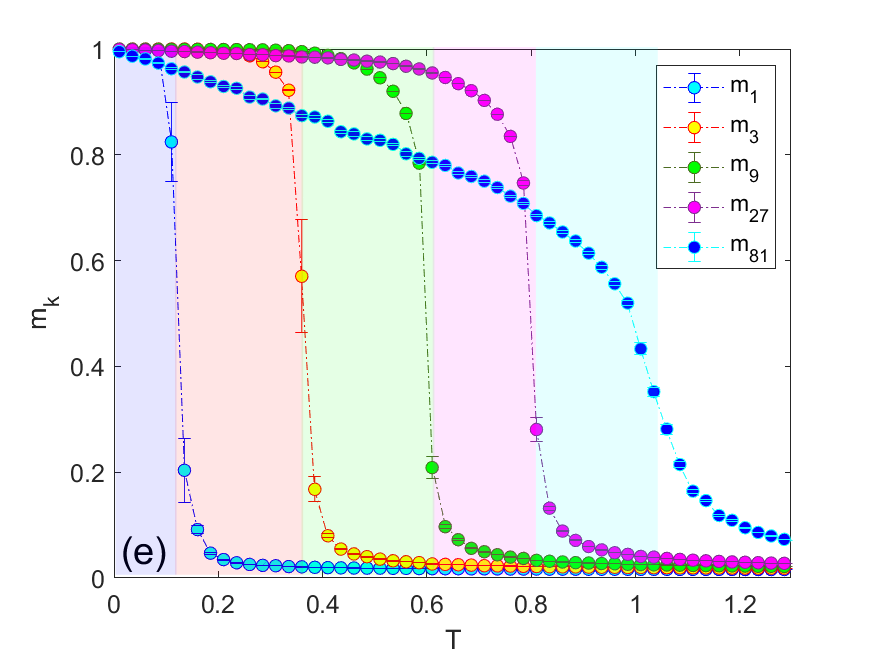}\label{fig:mk-T_q3_k5_Jincr}}\hspace*{-2mm}
\subfigure{\includegraphics[scale=0.4,clip]{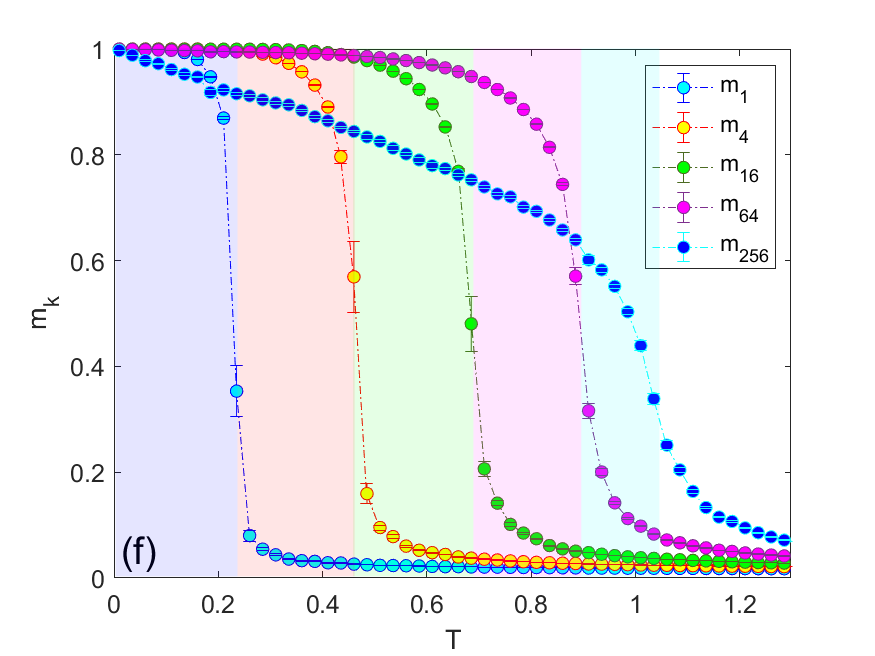}\label{fig:mk-T_q4_k5_Jincr}}\\
\subfigure{\includegraphics[scale=0.4,clip]{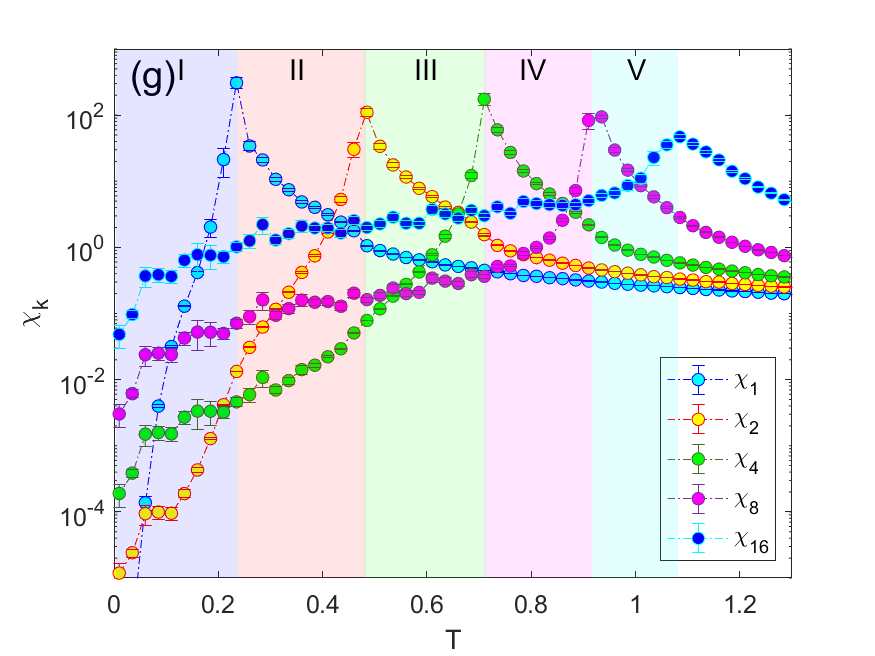}\label{fig:chik-T_q2_k5_Jincr}}\hspace*{-2mm}
\subfigure{\includegraphics[scale=0.4,clip]{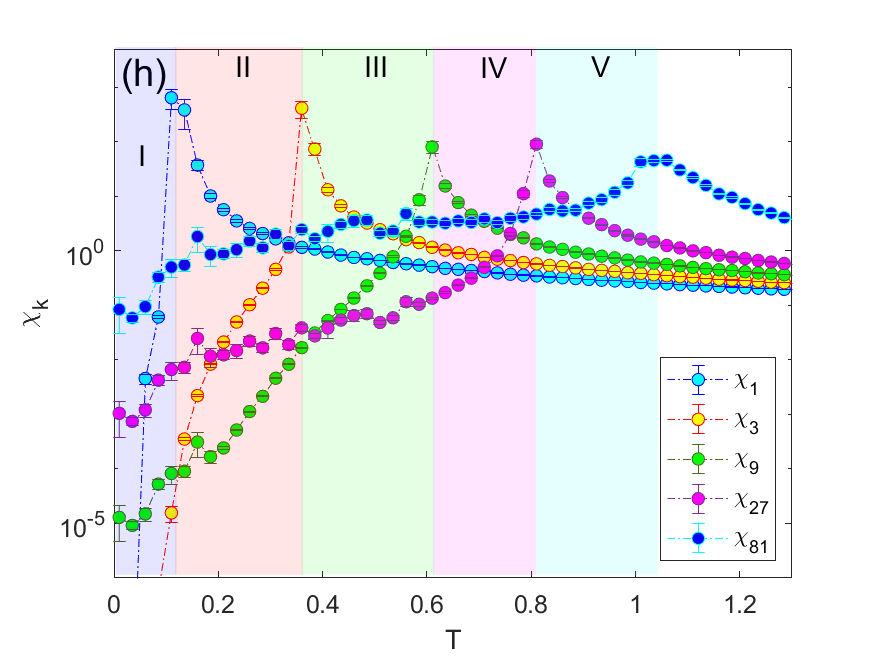}\label{fig:chik-T_q3_k5_Jincr}}\hspace*{-2mm}
\subfigure{\includegraphics[scale=0.4,clip]{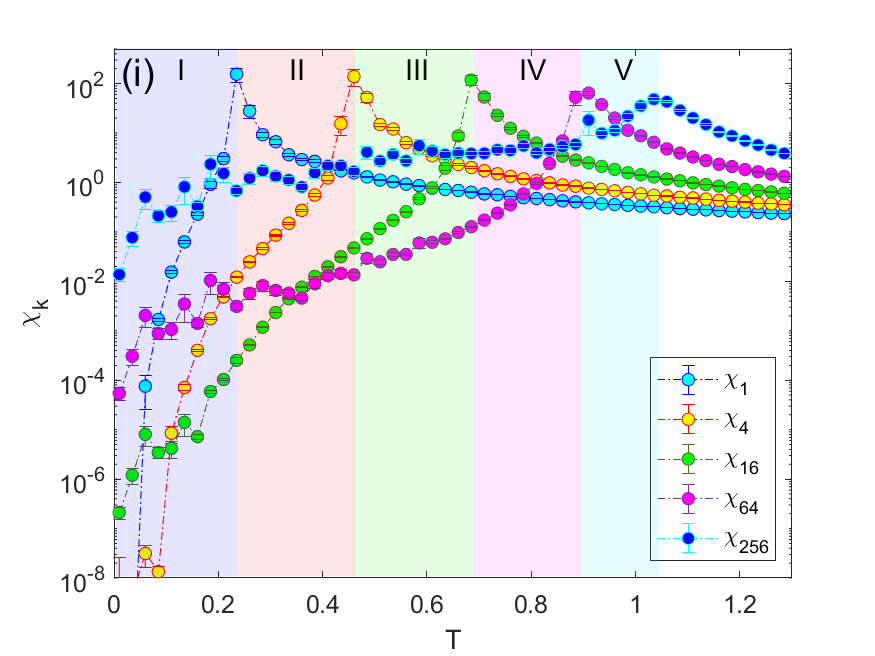}\label{fig:chik-T_q4_k5_Jincr}}
\caption{(Color online) Temperature dependencies of (a-c) the specific heat (d-f) the generalized magnetizations and (g-i) the generalized susceptibilities, for $n_t=5$ and (a,d,g) $q=2$, (b,e,h) $q=3$, and (c,f,i) $q=4$. The coupling constants are set to the values: $J_{1}=0.1$, $J_{2}=0.2$, $J_{4}=0.3$, $J_{8}=0.4$ and $J_{16}=1$ for $q=2$, $J_{1}=0.1$, $J_{3}=0.25$, $J_{9}=0.4$, $J_{27}=0.55$ and $J_{81}=1$ for $q=3$, and $J_{1}=0.2$, $J_{4}=0.4$, $J_{16}=0.6$, $J_{64}=0.8$ and $J_{256}=1$ for $q=4$. Background colors highlight approximate regions occupied by the phases I-V.}\label{fig:x-T_q2-4_nk5}
\end{figure} 

Now let us look at how this picture changes when the number of the terms in the Hamiltonian, $n_t$, is further increased. In Fig.~\ref{fig:x-T_q2-4_nk5} we present temperature dependencies of the calculated quantities for $n_t=5$ and again different values of $q$. For the cases of $q=2,3$ and $4$ one can observe overall five peaks in the specific heat curves, plotted in panels (a-c). Similar to the $n_t=3$ case, as the temperature decreases the indistinct round peak, that occurs for all $q$ at about the same temperature, is followed by another $n_t-1=4$ sharp peaks. The remaining panels (d-i) demonstrate that the respective peaks in the specific heat correspond to the five phase transitions $\rm{P \to V \to IV \to III \to II \to I}$, where the ordered phases I-V, characterized by the decay of the order parameters $m_{1},m_{q},\hdots,m_{q^{n_t-1}}$, are denoted in panels (g-i).

\begin{figure}[t!]
\centering
\subfigure{\includegraphics[scale=0.4,clip]{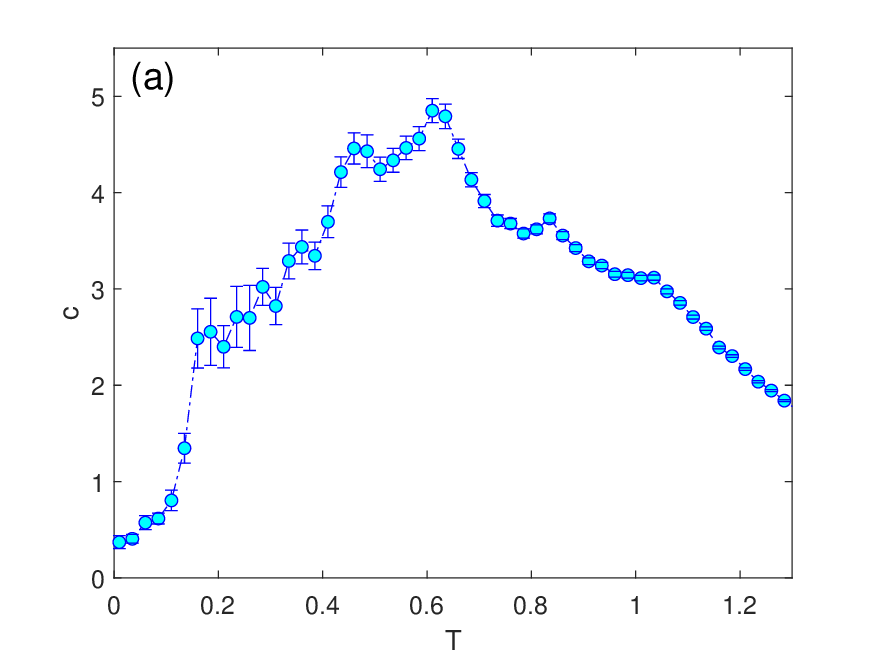}\label{fig:c-T_q5_k5_Jincr}}\\
\subfigure{\includegraphics[scale=0.4,clip]{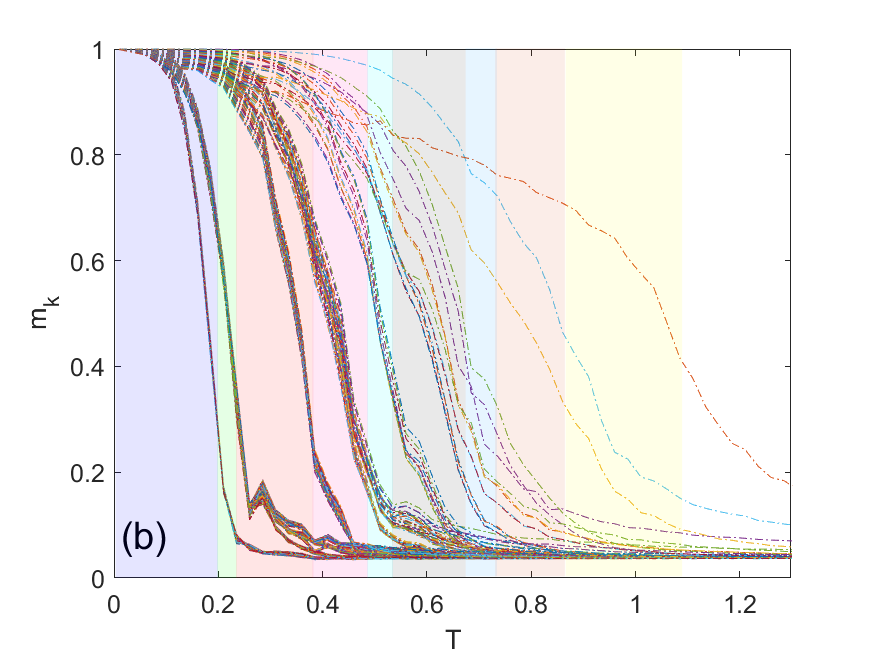}\label{fig:mk-T_q5_k5_Jincr}}\hspace*{-2mm}
\subfigure{\includegraphics[scale=0.4,clip]{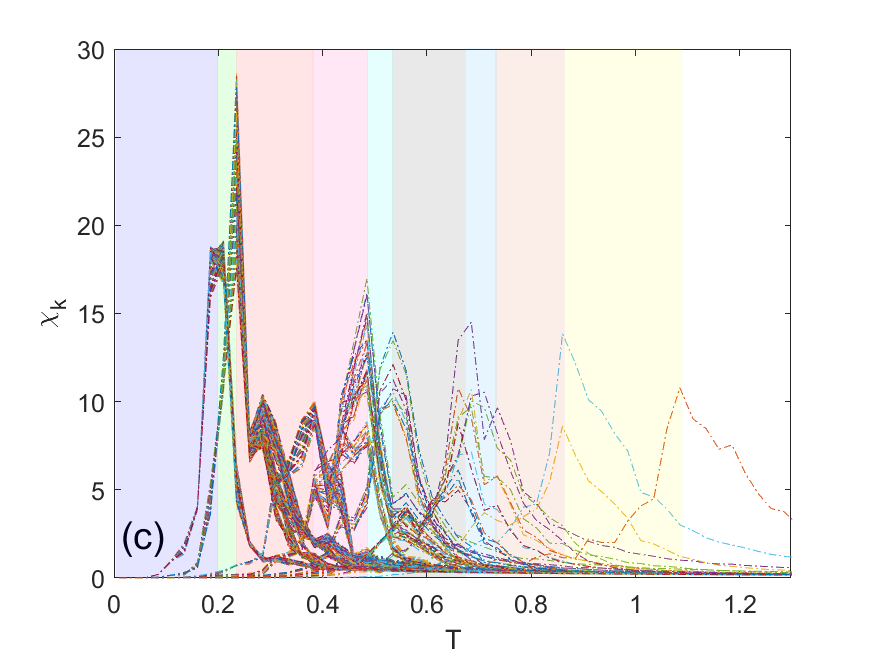}\label{fig:chik-T_q5_k5_Jincr}}\\
\subfigure{\includegraphics[scale=0.4,clip]{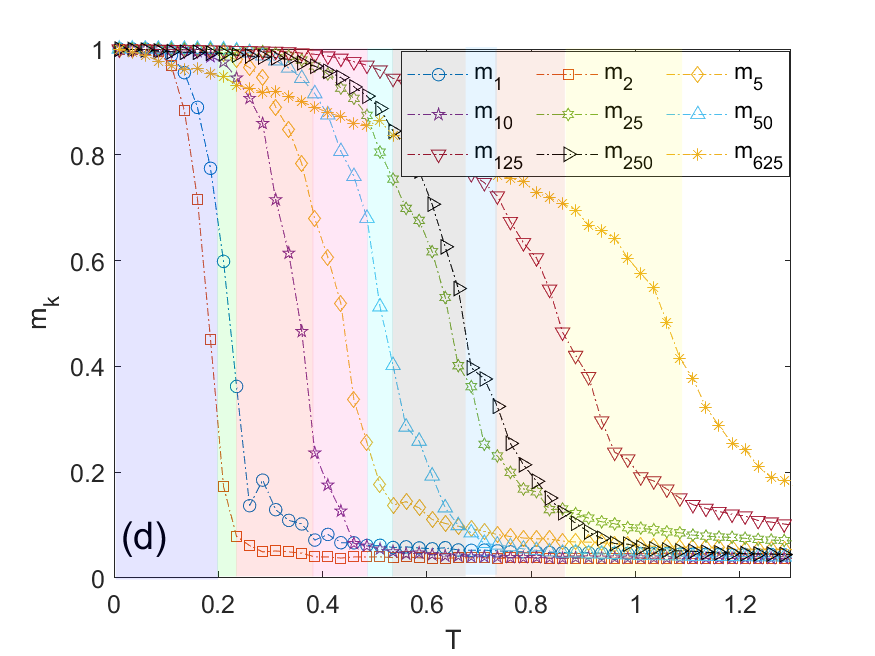}\label{fig:mk-T_q5_k5_Jincr_restr}}\hspace*{-2mm}
\subfigure{\includegraphics[scale=0.4,clip]{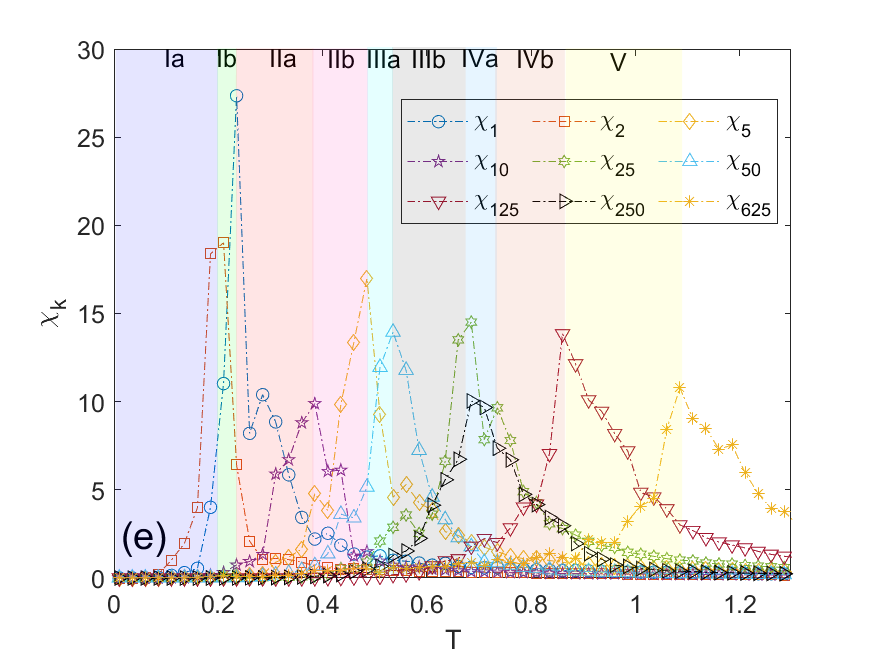}\label{fig:chik-T_q5_k5_Jincr_restr}}
\caption{(Color online) Temperature dependencies of (a) the specific heat (b) the generalized magnetizations and (c) generalized susceptibilities, for $q=5$ and $n_t=5$. Panels (d) and (e) show the generalized magnetizations and  generalized susceptibilities for the representative order parameters $m_k$. The coupling constants are set to the values: $J_{1}=0.2$, $J_{5}=0.4$, $J_{25}=0.6$, $J_{125}=0.8$ and $J_{625}=1$.}\label{fig:x-T_q5_nk5}
\end{figure}

Like in the case of $n_t=3$, the scenario of one BKT transition from the paramagnetic phase followed by apparently non-BKT $n_t-1$ phase transitions at lower temperatures changes for $q>4$. The evaluated quantities for $q=5$ are presented in Fig.~\ref{fig:x-T_q5_nk5}. The specific heat curve in panel (a) looks rather noisy (particularly at lower temperatures) with a number of anomalies but no sharp peaks. The plots of the quantities $m_k$ and $\chi_k$, $k=1,2,\hdots,5^{4}=625$ in panels (b) and (c) reveal the source of these anomalies. All the curves can be grouped to nine classes, based on the criterion that the values of $m_k$ decay and $\chi_k$ display peak at the same temperature. For clarity, the representative cases of $m_k$ and $\chi_k$, for $k=1,2,5,10,25,50,125,250$ and $625$, are presented in panels (d) and (e), respectively. These curves distinguish nine different ordered phases: Ia, Ib, IIa, IIb, IIIa, IIIb, IVa, IVb and V.

\begin{figure}[t!]
\centering
\subfigure{\includegraphics[scale=0.4,clip]{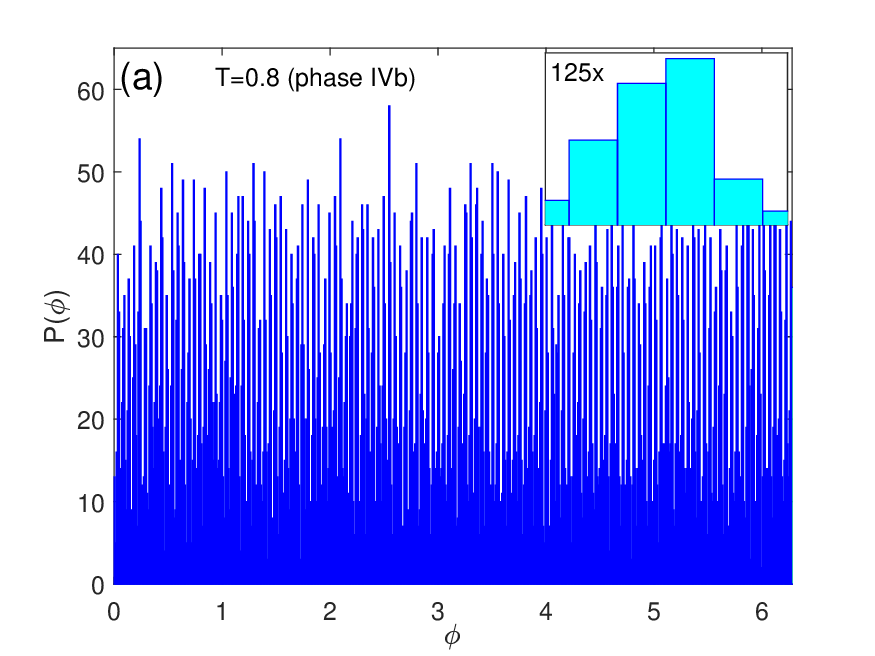}\label{fig:hst_q5_nk5_T0_8_L120}} \hspace*{-2mm}
\subfigure{\includegraphics[scale=0.4,clip]{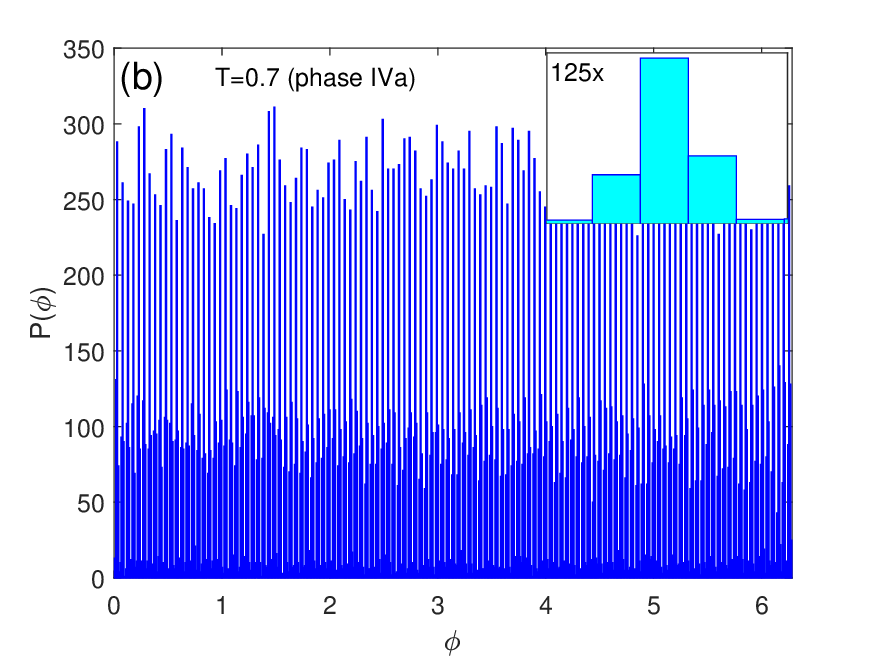}\label{fig:hst_q5_nk5_T0_7_L120}} \\
\subfigure{\includegraphics[scale=0.4,clip]{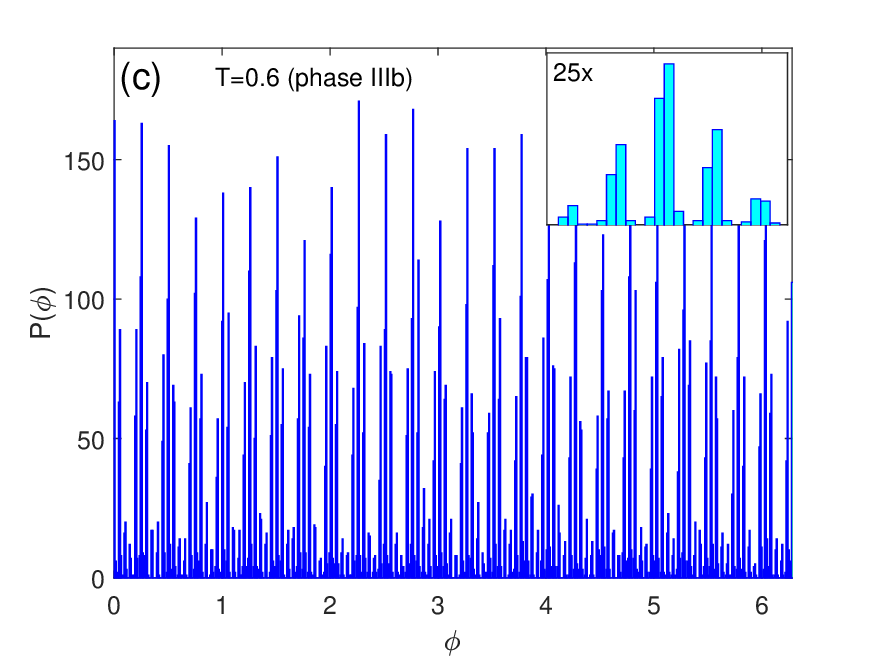}\label{fig:hst_q5_nk5_T0_6_L120}} \hspace*{-2mm}
\subfigure{\includegraphics[scale=0.4,clip]{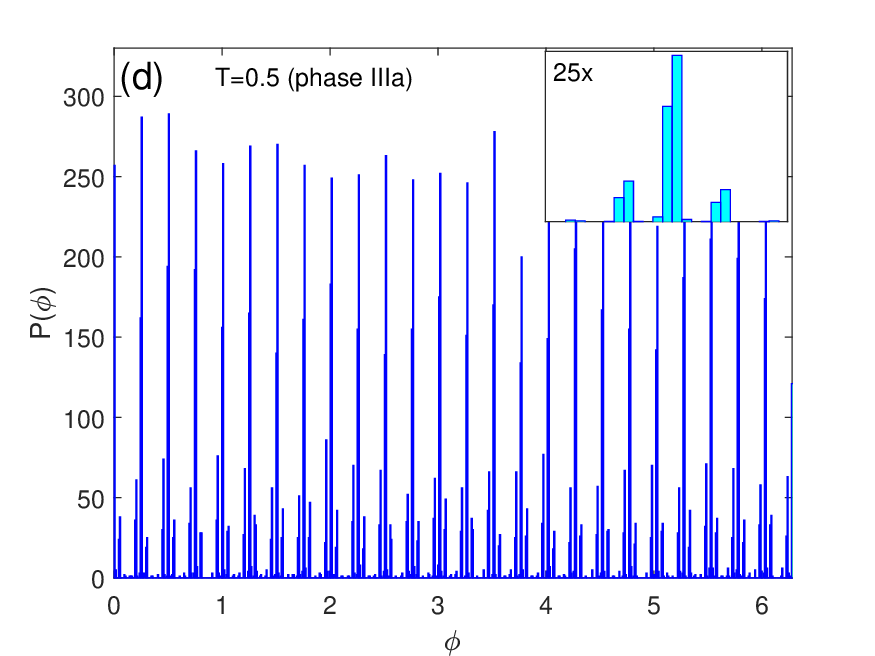}\label{fig:hst_q5_nk5_T0_5_L120}} \\
\subfigure{\includegraphics[scale=0.4,clip]{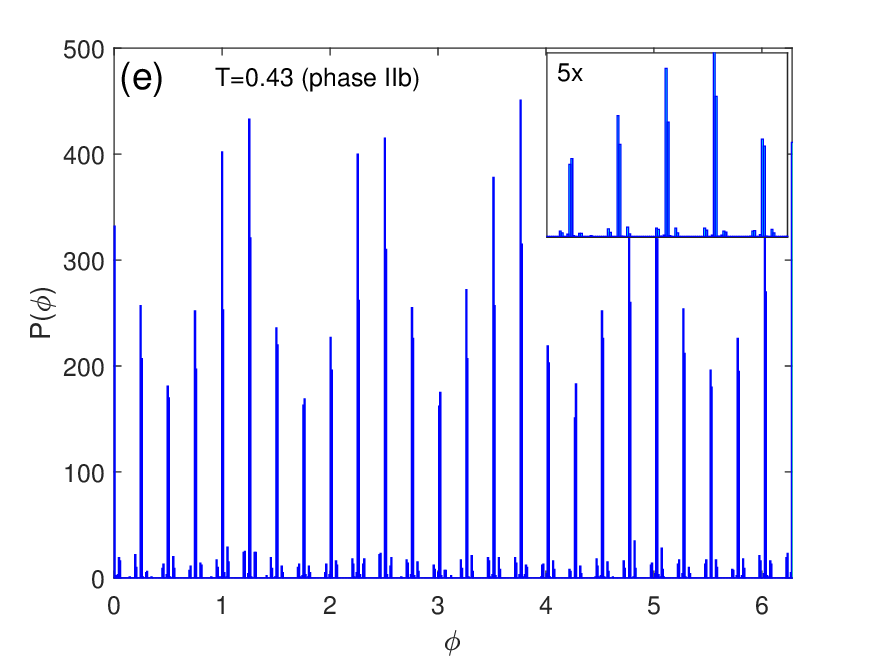}\label{fig:hst_q5_nk5_T0_43_L120}} \hspace*{-2mm}
\subfigure{\includegraphics[scale=0.4,clip]{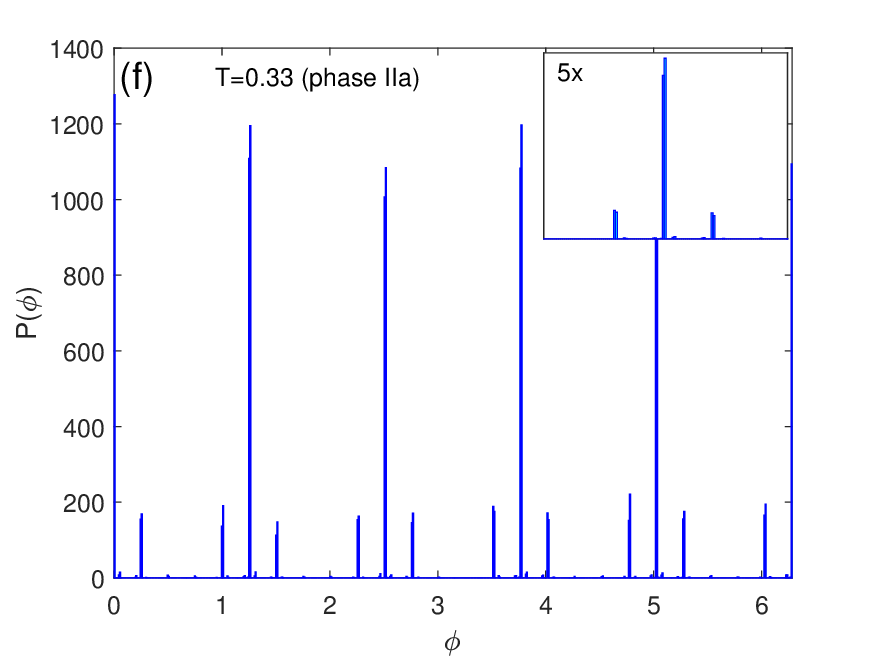}\label{fig:hst_q5_nk5_T0_33_L120}}\\
\subfigure{\includegraphics[scale=0.4,clip]{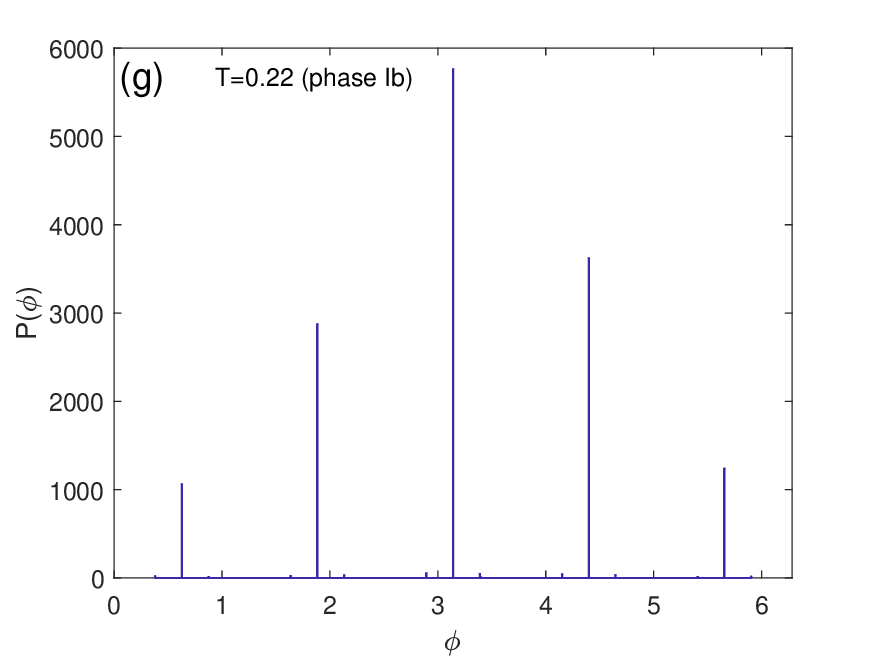}\label{fig:hst_q5_nk5_T0_22_L120}}
\caption{(Color online) Histograms at the representative temperatures corresponding to different phases, for $n_t=5$ and $q=5$.}\label{fig:hist_snap_nk5}
\end{figure} 

The nature of the identified phases can be again better understood by exploring spin distributions. The phases Ia and V have the same character as the phases I and V for $q<5$. The spin distributions in theses phases are rather trivial with all spins pointing in one direction (phase Ia) and $q^{n_t-1}=625$ equally populated preferential directions (phase V). The spin distributions in the remaining ordered phases are presented in Fig.~\ref{fig:hist_snap_nk5} with the bin width equal to $2\pi/625$. All the phases can be characterized by periodically repeating structures that are shown in detail in the insets. The histograms in the highest-symmetry ordered phase V can be attributed to the presence of the coupling $J_{625}$ and contains 625 bins of equal height\footnote{Due to the limited statistics on a finite-size lattice, there is some scatter of the heights.} (625x1 structure - not shown). In the phase IVb one can observe decomposition of this flat histogram into a periodic structure with 125 modes, each including 5 bins of different heights (125x5 structure). The emergence of this phase can be ascribed to the increasing effect of the coupling $J_{125}$. This coupling becomes dominant in the phase IVa, which retains 125 modes but in each mode two of the five states at the tails become practically completely suppressed. Here we note that even though the existence of the phase IVa is not so apparent in Fig.~\ref{fig:x-T_q5_nk5}, due to the rather noisy behavior of the curves scattered in a narrow temperature range, from the histograms it appears that the phase IVa emerges from IVb in a similar manner as the phase IIa from IIb for the $n_t=3$ case. Further decrease of temperature enhances the effect of still lower order terms ($J_{25}$) and the system transitions in the phase IIIb with only 25 modes.  As shown in the inset of panel (c), each mode includes 25 unevenly populated bins. Namely, each mode can be viewed as consisting of 5 smaller nodes of 5 bins with the most populated one in the center. This 25x25 structure persists also in the phase IIIa but two smallest nodes in each of the 25 larger nodes become suppressed. Notice this similarity of the IIIb-IIIa transition with the IVb-IVa transition at which two smallest bins in each of the 125 nodes become suppressed. Upon further decrease of temperature the dominance of $J_{5}$ shows up in the phase IIb by splitting all the bins into 5 large structures with 125 unevenly populated bins (5x125 structure - see the inset in panel (e)). The transition to the phase IIa can be again characterized by suppression of two of the smallest nodes in each of the 5 large nodes. Finally, the increasing effect of $J_1$ suppresses additional two nodes leaving only 5 very narrow unevenly populated nodes (see panel (g)), resulting in a kind of the ferromagnetic phase with nonvanishing magnetic moment. Full dominance of $J_1$ is realized at the lowest temperatures in the phase Ia, in which all the modes are suppressed except of the one that dominated in the phase Ib. 

Similar features in the critical behavior seem to persist also for larger number of the terms, $n_t$. In our simulations for $n_t=10$ and $q<5$ we could observe ten separate phase transitions, characterized by one round and nine sharp peaks in the specific heat, presumably related to one BKT and nine non-BKT phase transitions (see the results for $q=2$ in Ref.~\cite{zuko24}). However, with the increasing value of $n_t$ it becomes gradually more and more difficult to distinguish numerous phase transitions, particularly $2n_t-1$ transitions for $q=5$.

\subsection{\label{subsec:results_symmetry}Symmetry breaking and universality}

Below, let us briefly summarize the observed phase transitions and discus their types (universality) in the context of the symmetries of the individual phases and their possible breaking at the respective transitions. At high temperatures the model starts from a full continuous rotation symmetry (U(1) of the $XY$ spins). The paramagnetic to first ordered transition is topological (BKT): it does not break a continuous symmetry in the Landau sense but produces QLRO via vortex–antivortex binding. The Hamiltonian term $\cos(q^{n_t-1}\phi_{i,j})$ makes the energy invariant under rotations by $2\pi/q^{n_t-1}$. In this phase the continuous U(1) symmetry is \emph{effectively} reduced to a discrete cyclic symmetry $\mathbb{Z}_{q^{n_t-1}}$ (i.e. spins preferentially occupy $q^{n_t-1}$ directions). However, in 2D this manifests as QLRO rather than true LRO because the transition is BKT.											

All subsequent transitions at lower temperatures involve discrete reductions of the residual rotational symmetry (a $q$-fold reduction at each step). In particular, going from a nematic phase with $q^k$ preferred directions to one with $q^{k-1}$ directions reduces the residual discrete rotational symmetry by a factor of $q$: effectively $\mathbb{Z}_{q^k} \to \mathbb{Z}_{q^{k-1}}$. This is a discrete symmetry breaking (removal of $q$-fold degeneracy at that scale). As for the transition universality (between different nematic phases and to FM), for $q=2$, $3$ and $4$, those discrete symmetry-breaking transitions fall into familiar 2D universality classes (Ising for $q = 2$ and $4$; three-states Potts for $q = 3$). Namely, for $q = 2$ and $4$, the discrete reduction at each step is effectively a $\mathbb{Z}_2$ (up-down) type symmetry breaking and thus the Ising universality, while for $q = 3$, the discrete reduction is $\mathbb{Z}_3$, corresponding to the three-states Potts universality. It is interesting that the present continuous-variable models, due to the presence of the nematic terms, show a critical behavior that is very similar to that observed in a discrete $\mathbb{Z}_{q}$ symmetry Clock model.	

For $q = 5$, the intermediate phases split into two kinds, producing many more distinct phases with domain structures and unequal populations of preferred angles. The symmetry picture is no longer a simple successive $\mathbb{Z}_{q^k} \to \mathbb{Z}_{q^{k-1}}$ reduction in an obvious way because couplings of different orders compete and produce domain-structured states (short-range FM inside domains, phase shifts between domains). The transitions for $q = 5$ do not show clean agreement with standard 2D universality classes; in the FSS analysis some exponent estimates deviate, and one low-$T$ order parameter shows bimodal histograms (possible first-order or sluggish dynamics).

\section{\label{sec:concl}Conclusion}

Building upon the standard $XY$ model, we proposed generalized variants capable of exhibiting an arbitrary number of phase transitions. These models are constructed by supplementing the magnetic coupling with $n_t-1$ nematic terms of exponentially increasing order with the base $q=2,3,4$ and 5, and increasing interaction strength. If the order increases as the power of $q=2,3$ and 4 and the interaction strength of the final term is sufficiently large compared to the preceding one, the resulting model exhibits a number of phase transitions equal to the number of terms in the generalized Hamiltonian. Starting from high temperatures, the observed phases include $n_t-1$ nematic phases of the orders $q^{k}$, $k=n_t-1,n_t-2,\hdots,1$, that enforce $q^{k}$ preferential directions symmetrically disposed around the circle, followed by the ferromagnetic phase at the lowest temperatures. Instead of the BKT transition to the ferromagnetic phase in the standard $XY$ model, in these models this transition occurs between the paramagnetic and the highest-order nematic phase. All the remaining phase transitions at lower temperatures have a non-BKT nature: depending on the value of $q$ they belong to either the Ising ($q=2$ and $4$) or the three-states Potts ($q=3$) universality class.

For $q=5$ the number of the observed phase transitions is even larger than the number of terms in the Hamiltonian. It appears that, except for the the highest-order nematic phase, all the remaining phases split into two, resulting in overall $2 n_t$ phases, including the paramagnetic one. The new phases that are not observed for $q<5$ emerge due to the interplay between different terms. As temperature decreases the effect of the lower-order terms increases, which results in a series of different phases characterized by the gradually decreasing preferential directions that may be unequally represented. Consequently, above the standard ferromagnetic phase, there is another FM phase with $q$ unequally populated directions around the circle and thus non-zero magnetic moment and an algebraically decaying spin correlation function. The characteristic spin configurations have domain structures with relatively large domains with FM arrangement of the spins inside the domains but the phase shift $2\pi/q$ among domains. In all the nematic phases the spin correlation function decays exponentially but, owing to the increasing effect of the magnetic interaction with the decreasing temperature, the correlation length gradually increases.

Apparently, one cannot overlook some similarity with the critical behavior of the discrete Clock model, as far as the number of the transitions between different QLRO phases and their nature is concerned. Namely, for $q<5$ the transition from the para to the FM LRO phase in the $q$-state Clock model is of the same universality as the transitions between any pair of the neighboring QLRO phases in the present model: the Ising universality class for $q = 2$ and $4$, and the three-states Potts model for $q = 3$~\cite{jose77,elit79}. For $q \geq 5$, in the Clock model there is a quasiliquid intermediate phase emerging between the high-temperature paramagnetic and the low-temperature FM phase, separated by two phase transitions. For $q \geq 8$ the high-temperature transition belongs to the BKT universality class but for $q =5 $ both transitions have been reported to be different from BKT~\cite{lapi06,baek13}. In the present models, for $q=5$ each transition between the QLRO phases also splits into two separate transitions, which do not seem compatible with the BKT universality either. Further studies of the present models for still larger values of $q$ are required to verify if, like in the Clock model, the BKT universality is eventually restored.

It is in order to state that the observed phenomena pertain to a very narrow region of the multidimensional model parameter space. Consequently, one should be precautious when generalizing the models' properties based on the obtained results. For example, we have found that if the strength of the last term is not sufficiently large then the number of the transitions in the cases with $q<5$ is reduced from $n_t$ to $n_t-1$ or if the order of the nematic terms increases not exponentially but linearly then even for the increasing interactions all the transitions bellow the BKT one can collapse to a single one~\cite{zuko24}. Additionally, if the coupling constants decay then there is just a single BKT transition that under certain circumstances can become first order~\cite{zuko17}. These results highlight the potential of the generalized $XY$ models as a versatile platform for investigating novel and unexpected critical phenomena.

A future extension of this work could involve a similar study for a 3D lattice, where one can expect a significantly different critical behavior. In particular, instead of the QLRO nature of the phases and the high-temperature BKT transition, such 3D systems are expected to show a true LRO with the transition to the paramagnetic state belonging to the 3D $XY$ universality class. Such a behavior was confirmed in a simpler $J_1-J_q$ $XY$ model in 3D~\cite{cano16} and the study also showed that the nature of the phase transitions at lower temperatures, including the whole phase diagram topology, may be different compared to the 2D case. Based on these results, there are indications that some of the transitions observed for $q>4$ in the present 2D models may disappear or change to crossovers in 3D cases. However, as pointed out in~\cite{cano16}, a reliable analysis of the critical behavior in such models may present a serious computational challenge requiring more sophisticated approaches.


\end{document}